\pdfoutput=1
\documentclass[journal]{IEEEtran}

\usepackage{amsmath,amssymb}
\usepackage{graphicx}
\usepackage{booktabs}
\usepackage{array}
\usepackage[affil-it]{authblk}
\usepackage[compress]{cite}

\usepackage{float} % for [H]
\usepackage{caption}
\usepackage{subcaption}
\graphicspath{{figs/}{figs/Binaural\space Error/}{figs/FOV/}{figs/FOV/Bin\space Error/}{figs/ILD/}{figs/ITD/}}

\usepackage[hidelinks]{hyperref}

\title{Binaural Signal Matching with Wearable Arrays for Near-Field Sources and Directional Focus}

\author[1]{Sapir Goldring}
\author[2]{Zamir Ben Hur}
\author[2]{David Lou Alon}
\author[2]{Chad McKell}
\author[2]{Sebastian Prepeli\c{t}\u{a}}
\author[1]{Boaz Rafaely}

\affil[1]{School of Electrical and Computer Engineering, Ben-Gurion University of the Negev}
\affil[2]{Reality Labs Research, Meta, Redmond, WA, USA}
\begin{document}

\maketitle

\begin{abstract}
This paper investigates the performance of Binaural Signal Matching (BSM) methods for near-field sound reproduction using a wearable glasses-mounted microphone array. BSM is a flexible, signal-independent approach for binaural rendering with arbitrary arrays, but its conventional formulation assumes far-field sources. In our previous work, we proposed a near-field extension of BSM (NF-BSM) that incorporates distance-dependent modeling and showed improved performance over far-field BSM using analytic data, though degradation persisted for sources very close to the array. In this study, we extend that analysis by using realistic simulated data of near-field Head-Related Transfer Functions (HRTFs) and Acoustic Transfer Functions (ATFs) of the array, accounting for listener head rotation and evaluating binaural cues such as interaural level and time differences (ILD and ITD). A key contribution is the introduction of a Field of View (FoV) weighting, designed to emphasize perceptually relevant directions and improve robustness under challenging conditions. Results from both simulation and a listening test confirm that NF-BSM outperforms traditional far-field BSM in near-field scenarios, and that the proposed NF-FoV-BSM method achieves the best perceptual and objective quality among all tested methods, particularly at close source distances and under head rotation. These findings highlight the limitations for far-field models in near-field sources and demonstrate that incorporating source distance and directional weighting can significantly improve binaural reproduction performance for wearable spatial audio systems.
\end{abstract}

% IEEE Keywords

% Introduction
% Introduction

\section{Introduction} \label{sec}

In recent years, applications such as Virtual Reality (VR) and teleconferencing have garnered significant interest due to their ability to provide immersive and interactive experiences \cite{madmoni2018direction, rafaely2022spatial, richard2023audio, ben2017spectral}. Central to these applications is the binaural reproduction of sound, which aims to recreate an acoustic scene using headphones, as it would be perceived by a listener. This process requires information about both the sound field and the Head-Related Transfer Functions (HRTFs) \cite{madmoni2025design}.

One approach to binaural reproduction is to incorporate recorded acoustic scenes represented by Ambisonics signals, which can be computed from spherical microphone arrays \cite{rafaely2015fundamentals}. Ambisonics has been widely studied for its ability to represent sound fields with high spatial accuracy \cite{rafaely2022spatial}, and can be directly convolved with HRTFs to compute binaural signals \cite{gerzon1985ambisonics}. However, while Ambisonics can be readily encoded using spherical arrays, encoding it with arrays of arbitrary configuration remains challenging \cite{rafaely2022spatial}.
To address this limitation, several methods have been proposed. One class of approaches is based on signal-independent processing. The beamforming-based binaural reproduction (BFBR) technique was introduced in \cite{song2008using}, where a set of beamformers is applied to the microphone signals, followed by HRTF filtering. A theoretical framework for optimizing this method's parameters was later presented in \cite{ifergan2022selection}. A recent followup to \cite{ifergan2022selection} is binaural signal matching (BSM) \cite{madmoni2025design, madmoni2024design,deppisch2021end,ahrens2021spherical}, a method noted for its accuracy and flexibility with various array geometries, including a wearable array. BSM is based on estimating binaural signals by minimizing the mean-squared error (MSE) between the array steering vectors and the HRTFs, using a linear formulation. This method has shown promising results in reproducing acoustic scenes with high fidelity.
Other approaches are signal-dependent. For example, the method described in \cite{mccormack2022parametric} encodes microphone array signals into Ambisonic signals using spatial filtering that adapts to the signal content. This technique decomposes the sound field into direct and reverberant components, allowing higher spatial resolution and broader frequency coverage, and is particularly suited for irregular arrays.
However, all these methods— signal independent and dependent approaches- share a significant limitation: they assume far-field plane wave sources and have not been specifically studied for near-field sources.

The far-field assumption is generally valid when the sound source is sufficiently distant from the array. Yet, in scenarios where the sound source is close to the array, such as in cases where a speaker is very close to a person with a wearable array, the own voice of the person wearing the array, or when playing a musical instrument, the far-field assumption may not hold. Instead, the sound behaves more like a point source with a radial wavefront, making the plane wave approximation less accurate and leading to errors in the reproduction process \cite{fisher2010near}. In such near-field scenarios, the development of algorithms that can accurately reproduce the binaural signal remains an open problem.

Previous work by the authors \cite{goldring2025binauralsignalmatchingwearable} presented the first investigation of BSM in near-field conditions using analytic data derived from rigid-sphere models. This study showed that while incorporating source distance information into the BSM framework (near-field BSM, or in short NF-BSM) substantially reduced binaural error compared to traditional far-field BSM (FF-BSM), performance for very close sources remained limited.  

Building on that foundation, the present paper uses realistic numerically simulated near-field HRTFs and steering data to conduct a more comprehensive evaluation of BSM in near-field scenarios. In addition to refining the NF-BSM framework by adopting a mixed error formulation for binaural reproduction, the study examines perceptual outcomes through a listening test and analyzes binaural cues (ITD and ILD) under head rotations. Furthermore, this paper introduces a Field of View (FoV) weighting approach designed to improve reproduction accuracy by prioritizing directions of interest.

The main contributions of this paper are as follows:
\begin{itemize}
    \item A more realistic simulation study using realistic Finite Difference Time Domain (FDTD)-generated HRTFs and steering matrices tailored for wearable glasses-mounted arrays.
    \item An extended NF-BSM framework evaluated under head rotation and binaural cue metrics, with validation through a formal listening test.
    \item The introduction of a novel FoV weighting design for BSM in near-field conditions, demonstrating improved performance for relevant spatial regions.
\end{itemize}

\section{Mathematical background}
\label{MATH}
This section introduces the signal model used for microphone array processing, the binaural signal representation, and the framework for BSM, which form the foundation for understanding the proposed approach to near-field binaural reproduction.

\subsection{Microphone Array Measurement Model}

This subsection outlines the mathematical framework for modeling signals captured by a microphone array in the context of this study. A standard spherical coordinate system is adopted, represented by \((r, \theta, \phi)\), where \(r\) is the radial distance from the origin, \(\theta\) is the elevation angle measured from the Cartesian \(z\)-axis downward toward the \(xy\)-plane, and \(\phi\) is the azimuth angle measured from the positive \(x\)-axis in the direction of the positive \(y\)-axis. The wave number \(k\) is defined as \(k = \frac{2\pi}{\lambda}\), where \(\lambda\) is the wavelength, and \(f\) represents the frequency. Let \(\Omega = (\theta, \phi)\) denote the direction of arrival of a sound source in the sound  \cite{rafaely2015fundamentals}.

The sound field is assumed to be composed of \(Q\) sound sources, each with source signals \(s_q(k)\) assigned to the sources arriving from directions \(\{\Omega_q\}_{q=1}^Q\). This sound field is captured by an \(M\)-element microphone array centered at the origin. The recorded signals can then be represented by the following narrowband model \cite{madmoni2024design}:
\begin{equation}
    \mathbf{x}(k) = \mathbf{V}(k,\Omega)\,\mathbf{s}(k) + \mathbf{n}(k),
\label{eq:signal_model}
\end{equation}
\noindent where \(\mathbf{x}(k)=[x_1(k),\ldots,x_M(k)]^{\mathsf T}\) is the \(M\)-microphone signal vector; \(\mathbf{V}(k,\Omega)\in\mathbb{C}^{M\times Q}\) is the steering matrix whose \(q\)-th column is \(\mathbf{v}_q(k,\Omega_q)\); \(\mathbf{s}(k)=[s_1(k),\ldots,s_Q(k)]^{\mathsf T}\) is the source vector; \(\mathbf{n}(k)=[n_1(k),\ldots,n_M(k)]^{\mathsf T}\) is additive noise; and \(\Omega=\{\Omega_q\}_{q=1}^Q\). Superscript \({}^{\mathsf T}\) denotes transpose.

The steering vectors can be derived analytically or numerically for specific array types  \cite{rafaely2015fundamentals}, or obtained through direct measurement. The representation of these steering vectors for different source types, including plane waves and point sources, will be demonstrated in the simulation study.

\subsection{Binaural Signal Model}

Assuming that the listener’s head is positioned at the origin, the binaural signal can be expressed as:

\begin{equation}
p^{l/r}(k) = [\mathbf{h}^{l/r}(k)]^T \mathbf{s}(k)
\label{eq:reproduction}
\end{equation}%
where \(p^{l/r}(k)\) represents the binaural signal for the left (\(l\)) and right (\(r\)) ears, \([\mathbf{h}^{l/r}(k)]^T\) denotes the transposed HRTF vector for the left and right ears at wave number \(k\), and \(\mathbf{s}(k)\) is the source signal vector.

The HRTF vector \([\mathbf{h}^{l/r}(k)]^T\) can be defined as:

\begin{equation}
[\mathbf{h}^{l/r}(k)]= \begin{bmatrix} h^{l/r}(k, \Omega_1) & h^{l/r}(k, \Omega_2) & \dots & h^{l/r}(k, \Omega_Q) \end{bmatrix}^T
\end{equation}%
where \(h^{l/r}(k, \Omega_q)\) represents the HRTF from the acoustic source located at direction \(\Omega_q\) to the left or right ear of the listener.

It is important to note that HRTFs vary with source distance (e.g., near-field vs far-field), which can influence the binaural signal measured at the listener’s ears \cite{kan2009psychophysical}.

\subsection{Binaural Signal Matching}

In the BSM approach, the binaural signal is estimated by applying weights to the microphone signals and summing them in a manner similar to beamforming. The estimated binaural signal can be expressed as:

\begin{equation}
\hat{p}^{l/r}(k) = [\mathbf{c}^{l/r}(k)]^H \mathbf{x}(k)
\label{estimated sig}
\end{equation}%
where \(\hat{p}^{l/r}(k)\) represents the estimated binaural signal for the left (\(l\)) and right (\(r\)) ears, \(\mathbf{c}^{l/r}(k)\) is an \(M \times 1\) complex vector of filter weights, \((\cdot)^H\) denotes the Hermitian (conjugate transpose) operator and \(\mathbf{s}(k)\) is the source signal vector.

The filter coefficients \(\mathbf{c}^{l/r}(k)\) are calculated as the solution to the minimization of the mean squared error \(\epsilon^{l/r}\) defined by:

\begin{equation}
\mathbf{c}^{l/r} = \arg \min_{\mathbf{c}} \, \epsilon^{l/r}
\end{equation}%
where the error \(\epsilon^{l/r}\) is given by:

\begin{equation}
\epsilon^{l/r} = \mathbb{E} \left[ \left| p^{l/r}(k) - \hat{p}^{l/r}(k) \right|^2 \right]
\end{equation}%
with \(\mathbb{E}[\cdot]\) representing the expectation operator.
The solution to this minimization problem, assuming the sound sources and the noise are i.i.d (independent and identically distributed), is given by:

\begin{equation}
\mathbf{c}^{l/r} = \left( \mathbf{V} \mathbf{V}^H + \frac{\sigma_n^2}{\sigma_s^2} \mathbf{I}_M \right)^{-1} \mathbf{V} [\mathbf{h}^{l/r}]^*
\label{weights}
\end{equation}%
Here, \(\sigma_n^2\) and \(\sigma_s^2\) are the noise and signal power respectively, \(\mathbf{I}_M\) is the \(M \times M\) identity matrix and \([\mathbf{h}^{l/r}]^*\) denotes the complex conjugate of the HRTF vector.

The BSM reproduction accuracy is measured by the normalized error, defined as:

\begin{equation}
\epsilon^{l/r}_{\text{LS}}(k) = \frac{\mathbb{E}\left[\left|p^{l/r}(k) - \hat{p}^{l/r}(k)\right|^2\right]}{\mathbb{E}\left[\left|p^{l/r}(k)\right|^2\right]}
\end{equation}%
This yields the following expression for the normalized error:

\begin{equation}
\epsilon^{l/r}_{\text{LS}}(k) = \frac{\sigma_s^2 \left\| \mathbf{V}^T \left(\mathbf{c}^{l/r}_{\text{LS}}\right)^* - \mathbf{h}^{l/r} \right\|_2^2 + \sigma_n^2 \left\| \left(\mathbf{c}^{l/r}_{\text{LS}}\right)^* \right\|_2^2}{\sigma_s^2 \left\| \mathbf{h}^{l/r} \right\|_2^2}
\label{error}
\end{equation}%
Here, \(\|\cdot\|_2\) denotes the \(L_2\)-norm.
For frequencies above approximately 1500~Hz, it has been shown that the perceptual interaural level difference (ILD) becomes more dominant than the interaural time difference (ITD) \cite{madmoni2024design}. Consequently, at high frequencies, the binaural signal matching criterion based on the MSE of the complex signal is replaced by an alternative criterion that matches the magnitude of the binaural signals instead.

This leads to the definition of the magnitude-based normalized error:

\begin{equation}
\epsilon^{l/r}_{\text{MagLS}}(k) = \frac{\left\| \left| \mathbf{V}^T \left(\mathbf{c}^{l/r}_{\text{MagLS}}\right)^* \right| - \left| \mathbf{h}^{l/r} \right| \right\|_2^2}{\left\| \left| \mathbf{h}^{l/r} \right| \right\|_2^2}
\label{magls_error}
\end{equation}
Here, \(\mathbf{c}^{l/r}_{\text{MagLS}}\) denotes the filter weights optimized to minimize this magnitude-based error.

Another error is defined that is composed of a smooth transition between the complex-valued error (important at low frequencies) and the magnitude-based error (important at high frequencies), denoted as mixed error criterion:

\begin{equation}
\epsilon^{l/r}_{\text{mix}}(k) = \alpha(k) \cdot \epsilon^{l/r}_{\text{LS}}(k) + [1 - \alpha(k)] \cdot \epsilon^{l/r}_{\text{MagLS}}(k)
\label{mixed_error}
\end{equation}
where \(\epsilon^{l/r}_{\text{LS}}(k)\) is defined as in Eq.~\eqref{error}, using the optimal weights \(\mathbf{c}^{l/r}_{\text{LS}}\), and \(\epsilon^{l/r}_{\text{MagLS}}(k)\) is computed using the corresponding optimal weights \(\mathbf{c}^{l/r}_{\text{MagLS}}\).

The weighting function \(\alpha(k)\) is defined as:

\begin{equation}
\alpha(k) =
\begin{cases}
1, & f_k < 800~\mathrm{Hz},\\[2pt]
\dfrac{1500 - f_k}{700}, & 800~\mathrm{Hz} \le f_k \le 1500~\mathrm{Hz},\\[2pt]
0, & f_k > 1500~\mathrm{Hz}.
\end{cases}
\label{eq:alpha-weight}
\end{equation}

where \(f_k\) is the frequency corresponding to index \(k\). This definition ensures that the error measure transitions smoothly from complex-valued matching at low frequencies to magnitude-based matching at high frequencies.

\section{Near-Field BSM: Mathematical Framework and Modeling}
\label{NF BSM}
This chapter presents the mathematical framework for applying the BSM algorithm to near-field sources \cite{goldring2025binauralsignalmatchingwearable}. The BSM formulation for near-field sources is introduced, showing the modifications required for steering functions and HRTFs to account for near-field effects. Two evaluation approaches for BSM are then described: one using filter coefficients designed assuming far-field sources but applied to near-field sources and the other using filters designed assuming near-field sources. 

\subsection{BSM Formulation for Near-Field Sources}

Assume that the sound field is composed of \(Q\) point sources. In this case, the signal model specified in Eq. (\ref{eq:signal_model}) can be written as:

\begin{equation}
\mathbf{x}(k) = \mathbf{V}_{\text{nf}}(k, \Omega,r_s)\mathbf{s}(k) + \mathbf{n}(k),
\end{equation}
where \(\mathbf{V}_{\text{nf}}(k, \Omega,,r_s)\) is the point source steering matrix, with \(r_s\) denoting the source distance. Each \(\mathbf{v}_q(k, \Omega_q,r_q)\) column in this matrix represents the steering vector of a point source located at \((\Omega_q,r_q\)) relative to the \(M\)-element microphone array. The point source steering matrix differs from the plane wave steering matrix due to the radial wavefront of sound propagation in the near field, as opposed to a planar wavefront in the far field.

The true binaural signal, specified in Eq. (\eqref{eq:reproduction}), becomes:

\begin{equation}
p^{l/r}(k) = [\mathbf{h}^{l/r}_{\text{nf}}(k,r_s))]^T\mathbf{s}(k),
\end{equation}
where \(\mathbf{h}^{l/r}_{\mathrm{nf}}(k,r_s)\) denotes the near-field HRTF vector for the left/right ear. It differs from its far-field counterpart because the wavefront curvature is distance-dependent and interacts with the head–torso–pinnae anatomy. In addition, distance-related mechanisms modify the cues: (i) proximity effects increase head shadowing and overall level (via the \(1/r\) decay), yielding larger interaural level differences (ILDs) in the near field \cite{brungart1999auditory}; and (ii) geometric parallax changes monaural spectral cues because, as a source approaches the head along a radial path, the source-to-ear angle varies with distance (a condition that generally does not hold in the far field) \cite{brungart1999auditory, brungart2002near}. In practice, real emitters are not ideal point sources, so HRTFs can exhibit source dependence; in our simulations we assume ideal point sources to isolate distance-related effects.

The estimated binaural signal is calculated as detailed in Eq.~(\ref{estimated sig}), using BSM filter coefficients \(\mathbf{c}^{l/r}(k)\) that are computed based on two modeling assumptions: far-field and near-field propagation. In both cases, the filter coefficients are frequency-dependent and derived using different optimization strategies depending on the frequency range, as governed by the mixed error criterion defined in Eq.~(\ref{mixed_error}).

\iffalse
\textbf{Far-Field BSM Filters:}  
For the far-field configuration, the BSM filter coefficients are computed assuming a plane-wave propagation model. The steering matrix \(\mathbf{V}_{\text{ff}}\) corresponds to far-field sources (at a distance of 1.5~m), and the HRTF vector \([\mathbf{h}^{l/r}_{\text{ff}}]\) represents the corresponding far-field HRTFs. For frequencies below 800~Hz, the filters are derived by solving the analytical expression in Eq.~(\ref{weights}), resulting in:

\begin{equation}
\mathbf{c}^{l/r}_{\text{ff}}(k) = \left( \mathbf{V}_{\text{ff}} \mathbf{V}_{\text{ff}}^H + \frac{\sigma_n^2}{\sigma_s^2} \mathbf{I}_M \right)^{-1} \mathbf{V}_{\text{ff}} [\mathbf{h}^{l/r}_{\text{ff}}]^*.
\label{c_ff}
\end{equation}

For frequencies above 1500~Hz, the BSM filter is computed by minimizing the magnitude-based error defined in Eq.~(\ref{magls_error}), yielding \(\mathbf{c}^{l/r}_{\text{MagLS,ff}}(k)\), as described in \ref{MATH}.

In the transition band between 800~Hz and 1500~Hz, the two filters are linearly interpolated using the weighting function \(\alpha(k)\) defined in Eq.~(\ref{mixed_error}), to produce a frequency-dependent filter that balances complex and magnitude matching.
\fi

\textbf{Far-Field BSM Filters:}  
For the far-field configuration, the BSM filter coefficients are computed assuming a plane-wave propagation model. The steering matrix \(\mathbf{V}_{\text{ff}}\) corresponds to far-field sources (at a distance of 1.5~m), and the HRTF vector \([\mathbf{h}^{l/r}_{\text{ff}}]\) represents the corresponding far-field HRTFs. Two filters are computed: The LS filter \(\mathbf{c}^{l/r}_{\text{LS,ff}}\) is obtained using Eq.~(\ref{weights}), by substituting \(\mathbf{V} = \mathbf{V}_{\text{ff}}\) and \(\mathbf{h} = [\mathbf{h}^{l/r}_{\text{ff}}]\), and used in the complex-valued error criterion (Eq.~(\ref{error})).

The filter \(\mathbf{c}^{l/r}_{\text{MagLS,ff}}(k)\) is computed by minimizing the magnitude-based error defined in Eq.~(\ref{magls_error}). After computing these two filters, they are used in the calculation of the mixed error \(\epsilon^{l/r}_{\text{mix}}\) detailed in Eq.(\ref{mixed_error}), with the weighting function \(\alpha(k)\) from Eq.~(\ref{eq:alpha-weight}).

\textbf{Near-Field BSM Filters:}  
For the near-field configuration, the BSM filters incorporate a point-source propagation model. The steering matrix \(\mathbf{V}_{\text{nf}}\) models spherical wavefronts from sources at finite distances (\(< 1.5~\text{m}\)), and the HRTF vector \([\mathbf{h}^{l/r}_{\text{nf}}]\) corresponds to the near-field HRTFs. 

Two filters are computed: the LS filter \(\mathbf{c}^{l/r}_{\text{LS,nf}}\), obtained using Eq.~(\ref{weights}) with \(\mathbf{V} = \mathbf{V}_{\text{nf}}\) and \(\mathbf{h} = [\mathbf{h}^{l/r}_{\text{nf}}]\), and the magnitude-based filter \(\mathbf{c}^{l/r}_{\text{MagLS,nf}}\), minimizing the criterion in Eq.~(\ref{magls_error}). As in the far-field case, the mixed error \(\epsilon^{l/r}_{\text{mix}}\) is then formed by linearly combining the two errors using \(\alpha(k)\) (Eq.~(\ref{mixed_error})).

\medskip

\section{Near-field BSM Design with Limited Field of View}
\label{FOV Design}
In many practical scenarios, sources are not uniformly distributed across the full spatial domain but are instead concentrated within a limited angular region—for example, in front of the listener during conversation or interaction with a device. To model such directional priors, we introduce a design framework that incorporates a defined \textit{Field of View} (FoV) into the BSM formulation.

The FoV in this work is selected as the region within some aperture in both azimuth and elevation with respect to the frontal direction of the head. This region is intended to model the listener's primary visual region in realistic scenarios, such as face-to-face communication or interactive environments.

To incorporate this directional prior, we modify both the steering matrix and the HRTF vector used in the BSM formulation. Specifically, we apply a spatial weighting function \(w(\theta, \phi)\) to each direction of arrival \((\theta, \phi)\), where:

\[
w(\theta, \phi) =
\begin{cases}
1, & \text{if } (\theta, \phi) \in \text{FoV} \\
\beta, & \text{otherwise}
\end{cases}
\]
Here, \(0 \leq \beta < 1\) is a small positive constant introduced to avoid completely ignoring the contribution of out-of-FoV directions, which would otherwise lead to instability or excessively large errors in these directions. The value of \(\beta\) can be  chosen through a grid search as the optimal tradeoff between minimizing errors within the FoV and avoiding excessive errors outside the FoV.

The modified steering matrix and HRTF vector are then defined as:

\begin{align}
\tilde{\mathbf{V}} &= \mathbf{W} \mathbf{V} \label{eq:fov_steering} \\
\tilde{\mathbf{h}}^{l/r} &= \mathbf{W} \mathbf{h}^{l/r} \label{eq:fov_hrtf}
\end{align}
where \(\mathbf{W} = \text{diag}(w(\theta_1, \phi_1), \dots, w(\theta_Q, \phi_Q))\) is a diagonal matrix of weights applied per DOA, and \(Q\) is the number of sampled directions.

These weighted versions are then used in the BSM filter computation exactly as before, whether in the far-field or near-field model. The goal of this design is to determine whether restricting attention to a perceptually relevant region of space enhances the performance differences between the two models.

\section{Performance Measures}

In this section, the performance measures used to evaluate the accuracy of BSM reproduction are defined. The metrics include the normalized BSM error, ILD error, ITD error, and the effective spherical harmonics (SH) order. Each measure captures a different aspect of the reproduction quality.

\subsection{Binaural Error: Near-Field BSM}

As an initial study of the reproduction accuracy for near-field BSM, the normalized binaural signal matching MSE is defined in Eq.~\eqref{error}. Substituting the near-field model, the normalized MSE can be written as:

\begin{equation}
\epsilon^{l/r}_{\text{complex}}(k) = \frac{\sigma_s^2 \left\| \mathbf{V}_{\text{nf}}^\top \left(\mathbf{c}^{l/r}_{\text{BSM}}\right)^* - \mathbf{h}_{\text{nf}}^{l/r} \right\|_2^2 + \sigma_n^2 \left\| \left(\mathbf{c}^{l/r}_{\text{BSM}}\right)^* \right\|_2^2}{\sigma_s^2 \left\| \mathbf{h}_{\text{nf}}^{l/r} \right\|_2^2}
\label{nf_complex_error}
\end{equation}
where \(\mathbf{V}_{\text{nf}}\) is the near-field steering matrix, and \(\mathbf{h}_{\text{nf}}^{l/r}\) are the near-field HRTF vectors.

Following the same rationale described earlier, the near-field mixed error is defined by substituting the far-field quantities \(\mathbf{V}\) and \(\mathbf{h}^{l/r}\) with their near-field counterparts \(\mathbf{V}_{\text{nf}}\) and \(\mathbf{h}_{\text{nf}}^{l/r}\) in the expressions for both \(\epsilon^{l/r}_{\text{complex}}(k)\) and \(\epsilon^{l/r}_{\text{MagLS}}(k)\) defined in Eq.(\ref{magls_error}).
\subsection{Interaural Level Difference (ILD) Error}

The ILD error quantifies the deviation in interaural level differences between the reproduced and reference binaural signals. The ILD is estimated by processing the binaural signals through Equivalent Rectangular Bandwidth (ERB) filter banks, as outlined in \cite{xie2013head}. Specifically, the ILD in decibels for each filter band is calculated as:

\begin{equation}
ILD_f(f_c, \Omega) = 10 \log_{10} \frac{\sum_{f=0}^{f_c^{\text{max}}} \left| G(f, f_c) p^l(f) \right|^2}{\sum_{f=0}^{f_c^{\text{max}}} \left| G(f, f_c) p^r(f) \right|^2},
\label{ILD1}
\end{equation}
where \(G(f, f_c)\) represents the ERB filter with central frequency \(f_c\) at frequency \(f\), and \(f_c^{\text{max}}\) denotes the maximum frequency of this ERB filter. An average over 32 filter bands spanning the range \(1.5 - 20 \, \text{kHz}\), generated using the Auditory Toolbox \cite{slaney1998auditory}, is used to compute the ILD as:

\begin{equation}
ILD(\Omega) = \frac{1}{32} \sum_{f_c} ILD_f(f_c, \Omega)
\label{ILD2}
\end{equation}
The ILD error is then calculated as the absolute difference between the ILD of the reproduced and reference signals:

\begin{equation}
\epsilon_{\text{ILD}}(\Omega) = \frac{1}{32} \sum_{f_c} \left| ILD_{\text{rep}}(f_c, \Omega) - ILD_{\text{ref}}(f_c, \Omega) \right|
\label{ILD3}
\end{equation}
where \(\text{ILD}_{\text{ref}}(f_c, \Omega)\) represents the ILD of the reference binaural signal, and \(\text{ILD}_{\text{rep}}(f_c, \Omega)\) is the ILD estimated from the reproduced binaural signal.

\subsection{Interaural Time Difference (ITD) Error}

The ITD error measures deviations in the interaural time differences between the reproduced and reference binaural signals. ITD is estimated using the group delay of the binaural signals \cite{katz2014comparative}. The group delay in seconds for the left and right ears is calculated as:

\begin{equation}
\tau^{(\mathrm{s})}_{l/r}(f)
= \frac{1}{f_s}\,
\operatorname{Re}\!\left[
\frac{\operatorname{DFT}\{\,n\,p^{l/r}(n)\,\}}
     {\operatorname{DFT}\{\,p^{l/r}(n)\,\}}
\right]
\label{Tau}
\end{equation}
where \(p^{l/r}(n)\) is the binaural signal for the left (\(l\)) or right (\(r\)) ear (either true or estimated), \(\operatorname{DFT}\{\cdot\}\) denotes the discrete Fourier transform, and \(f_s\) is the sampling rate (Hz). The ITD is then obtained by subtracting the group delay of the right ear from that of the left ear:

\begin{equation}
ITD(f) = \tau_l(f) - \tau_r(f).
\label{itd_est}
\end{equation}
To evaluate the ITD error, the ITD is averaged over frequencies up to \(1.5 \, \text{kHz}\), and the error is computed as the absolute difference between the reference and estimated ITD:

\begin{equation}
\epsilon_{\text{ITD}}(r) = \left| \text{ITD}_{\text{ref}}(r) - \text{ITD}_{\text{rep}}(r) \right|.
\label{ITDcalcc}
\end{equation}
\iffalse
\subsection{Effective Spherical Harmonics (SH) Order}

The effective SH order characterizes the spatial complexity of a sound field as characterized by the steering function or the HRTF. For the given functions represented in the SH domain, the overall variance up to a given order \(N\) is calculated as:

\begin{equation}
E(N) = \sum_{n=0}^{N} \sum_{m=-n}^{n} \left| f_{nm} \right|^2,
\end{equation}
where \(f_{nm}\) represents the SH coefficients of the given steering function or HRTF. The cumulative variance is normalized by dividing it by the total variance across all orders \(N_{\text{max}}\), which represents the maximum SH order used in the simulation:

\begin{equation}
\bar{E}(N) = \frac{E(N)}{\sum_{n=0}^{N_{\text{max}}} \sum_{m=-n}^{n} \left| f_{nm} \right|^2}.
\end{equation}
The effective SH order for each frequency is then defined as:

\begin{equation}
N_{\text{eff}} = \min \{ N : \bar{E}(N) > 0.95 \},
\label{effectiveOrder}
\end{equation}
where \(N_{\text{eff}}\) is the smallest SH order \(N\) such that the cumulative normalized energy \(\bar{E}(N)\) exceeds \(95\%\).
\fi

\subsection{Null Space Projection Measure}

The null space projection measure, introduced in \cite{gayer2024ambisonics}, quantifies the degree to which a vector resides within the null space of a matrix, or, equivalently, its deviation from the column space of the matrix.
In the context of this work, this measure is applied to determine whether the HRTF vector, \ \(\mathbf{h}_{\text{nf}}^{l/r}\), can be spanned by the column space of the steering matrix \(\mathbf{V}^\top\). The null space projection measure is formally defined as:

\begin{equation}
\xi_{\text{null}} = 10 \log_{10} \left( \frac{\left\| \mathbf{V}_0^\top \mathbf{h}_{\text{nf}}^{l/r} \right\|_2^2}{\left\| \mathbf{h}_{\text{nf}}^{l/r} \right\|_2^2} \right),
\label{null_measure}
\end{equation}
where the null space of matrix \(\mathbf{V}^\top\)  denoted by \(\mathbf{V}_0^\top\) is obtained from the singular value decomposition (SVD) of \(\mathbf{V}\). Specifically, \(\mathbf{V}_0^\top\) comprises the eigenvectors corresponding to singular values below a threshold of \(-20 \, \text{dB}\) relative to the largest singular vector, representing an approximation of the null space of \(\mathbf{V}\).

With this definition, lower values of the measure \(\xi_{\text{null}}\) reflect a stronger alignment of \(\mathbf{h}\) with the column space, thereby facilitating a more accurate reconstruction of the binaural signals. 
\iffalse
\subsection{Effective Rank Measure}
The effective rank, introduced in \cite{tourbabin2014theoretical} for microphone array analysis, serves as a refined measure of the information content of a matrix. Unlike the traditional rank, which simply counts the number of nonzero singular values, the effective rank evaluates the distribution of singular values using Shannon entropy, providing a continuous assessment of the matrix’s dimensionality.

The effective rank of a matrix \(\mathbf{H}\) is defined as:

\begin{equation}
R(\mathbf{H}) = \exp\left(-\sum_{i=1}^{q} \tilde{\sigma}_i \cdot \log \tilde{\sigma}_i \right),
\label{effective_rank}
\end{equation}
where \(\tilde{\sigma}_i\) are the normalized singular values of \(\mathbf{H}\), computed as:

\begin{equation}
\tilde{\sigma}_i = \frac{\sigma_i}{\sum_{j=1}^{q} \sigma_j}.
\end{equation}

Here, \(\sigma_i\) represents the singular values of \(\mathbf{H}\), arranged in descending order, and \(q = \text{rank}(\mathbf{H})\) is the traditional rank of the matrix. 

This measure represents the effective number of independent singular vectors in the matrix. A higher effective rank implies that more of the singular vectors are independent, indicating greater information capacity within the matrix. 
\fi

\section{Simulation Study}
This section presents a simulation study aimed at objectively quantifying the performance of the BSM algorithm for near-field sources. Unlike the analytical modeling approach used in our earlier work \cite{goldring2025binauralsignalmatchingwearable}, the present study relies on numerically simulated near-field data generated with the HATS manikin model, for both HRTFs and array steering functions. Therefore, this evaluation is more realistic compared to the sphere-based models.

The structure of this chapter is as follows: Section~\ref{sec:setup} presents the simulation setup, Section~\ref{sec:Methods} describes the methods of the simulation, Section~\ref{Results} reports the main findings, and Section~\ref{FoV} discusses the results of the Field-of-View (FoV) analysis.

\subsection{Setup}
\label{sec:setup}

To generate realistic near-field steering matrices and HRTFs, finite-difference time-domain (FDTD) simulations were performed, as detailed in \cite{gomez2020pinna}. The validation results in the cited work indicate that the simulated HRTFs agree with measured data within 1--2~dB up to 17~kHz, making them a reliable source for the present analysis. Our numerical simulation results are expected to extrapolate to HRTFs measured on humans because the boundary conditions of human skin are very similar to the rigid boundary conditions applied to the HATS surface in our simulations \cite{hajarolasvadi2024effect}. An ideal point source was employed, providing high omnidirectionality and strong time coherence - features critical for accurate near-field HRTF characterization. Six grids were used in the convergence study and asymptotic prediction, with the grid spacing $\Delta X$ varying from 0.6 mm to 0.96 mm.

The simulations also included a wearable array configuration, in which five microphones were mounted on a glasses frame positioned on the HATS manikin head \cite{oreinos2013measurement}. This allowed the extraction of both HRTFs and acoustic transfer functions (ATFs) of the array required for the BSM formulation. The geometry, source and receiver configurations, and solver parameters are outlined next.

\begin{figure}[ht]
  \centering
  \includegraphics[width=0.5\linewidth]{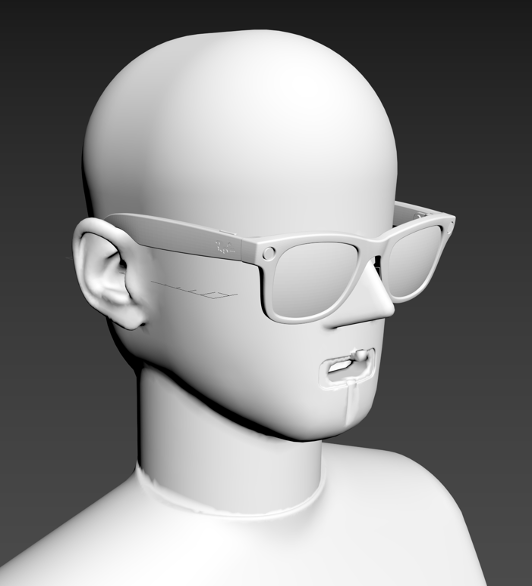}
  \caption{Glasses microphone array mounted on the HATS manikin.}
  \label{fig:glasses_on_hats}
\end{figure}

A Brüel \& Kjær Type 5128 Head-and-Torso Simulator (HATS) \cite{oreinos2013measurement} was used as both the measurement and simulation geometry. The HATS mesh used in the simulations was an accurate (error $\le 1$ mm) 3D scan of the HATS. Figure~\ref{fig:glasses_on_hats} shows the glasses array mounted on the HATS head.

Five microphones were integrated into the glasses frame at predefined locations. Their Cartesian coordinates are listed in Table~\ref{tab:mic_coords}. No special microphone porting was done: the pressure was acquired on the surface of the glasses.
\iffalse
\begin{table}[ht]
\small
  \centering
  \caption{Microphone positions on the glasses mesh. Coordinates are given in millimeters relative to the head center, with +x forward, +y to the left, and +z upward.}
  \label{tab:two_col_sources}
  \hspace*{-0cm}
  \begin{tabular}{|
      >{\centering\arraybackslash}m{0.30\linewidth}|
      >{\raggedright\arraybackslash}m{0.65\linewidth}|
    }
    \hline
    \textbf{Visualization} & \textbf{Microphone coordinates [mm]} \\
    \hline
    \includegraphics[width=\linewidth]{figs/Noise_Mic.png} & (100.774,\,-16.528,\,-5.079) \\
    \hline
    \includegraphics[width=\linewidth]{figs/Left_Mid_Tenple.png} & (31.093,\,76.620,\,20.619) \\
    \hline
    \includegraphics[width=\linewidth]{figs/Right_Mid_temple.png} & (30.894,\,-76.636,\,20.608) \\
    \hline
    \includegraphics[width=\linewidth]{figs/Left_Logo.png} & (85.988,\,73.327,\,28.807) \\
    \hline
    \includegraphics[width=\linewidth]{figs/Right_logo.png} & (86.327,\,-73.328,\,28.876) \\
    \hline
  \end{tabular}
\end{table}
\fi

\begin{table}[ht]
\small  % use smaller font for space efficiency
\centering
\caption{Microphone coordinates on the glasses frame, rounded to the nearest mm, relative to the head center.}
\label{tab:mic_coords}  % this label can be referenced in text using \ref{tab:mic_coords}
\begin{tabular}{|l|r r r|}  % l = left-align label; r = right-align numbers
  \hline
  \textbf{Label} & \textbf{x} & \textbf{y} & \textbf{z} \\
  \hline
  Nose              & 101 & $-$17 & $-$5 \\
  Left mid-temple   & 31  & 77    & 21 \\
  Right mid-temple  & 31  & $-$77 & 21 \\
  Left logo         & 86  & 73    & 29 \\
  Right logo        & 86  & $-$73 & 29 \\
  \hline
\end{tabular}
\end{table}

Finally, the point sources were positioned on a spherical grid around the HATS head at 2702 directions on a lebedev grid (azimuth–elevation pairs) and at six radial distances  
\[
  r_s \;=\;\{0.15,\,0.20,\,0.45,\,0.70,\,1.00,\,1.50\}\,\mathrm{m}
\]
from the head center. In contrast to the dataset used in our previous research \cite{goldring2025binauralsignalmatchingwearable}, where the far-field distance was 3.2~m, the far-field HRTF here is defined at 1.5~m due to simulation domain constraints. The ear canals of the HATS mesh were manually blocked such that blocked-meatus simulated HRTFs were used in this work. 
The Signal-to-Noise Ratio (SNR) was set to \(20 \, \text{dB}\) by appropriately choosing \(\sigma_s^2\) and \(\sigma_n^2\).

\subsection{Methods}
\label{sec:Methods}

In this study, performance was evaluated using two formulations of the BSM filter coefficients (\(\mathbf{c}_{\text{BSM}}^{l/r}\)), as detailed in the previous section: one using far-field BSM filters and the other using near-field BSM filters. 

To simulate head rotations of the listener during playback and analyze their influence on the BSM performance, we adopted a method similar to the one proposed in~\cite{madmoni2024design}. Specifically, the head rotation was modeled by modifying the azimuth angles of the HRTFs, assuming a rotation of the head in the horizontal plane. For a given rotation angle \(\Delta \phi\), the rotated HRTF vector \(\mathbf{h}_{\text{rot}}^{l/r}(k, r_s)\) is defined as:

\begin{equation}
\begin{aligned}
\mathbf{h}_{\text{rot}}^{l/r}(k, r_s) = \big[ & h^{l/r}(k, \theta_1, \phi_1 + \Delta \phi),\ 
                                               h^{l/r}(k, \theta_2, \phi_2 + \Delta \phi),\ \dots, \\
                                             & h^{l/r}(k, \theta_Q, \phi_Q + \Delta \phi) \big]^T
\end{aligned}
\end{equation}
Since only azimuthal rotations were considered, the elevation angles \(\theta_q\) were kept constant, and only the azimuth components \(\phi_q\) of the source directions were updated.

\subsection{Results: Analysis of Binaural Measures for Far-Field and Near-Field BSM}
\label{Results}
Performance was evaluated based on the measures detailed in Sec. IV, and is presented in this section.

\subsubsection{Binaural Error}
\begin{figure}[h!]
   \hspace{-0.2 cm}    
   \includegraphics[trim={45pt 0 0 0},clip,width=0.5\textwidth]{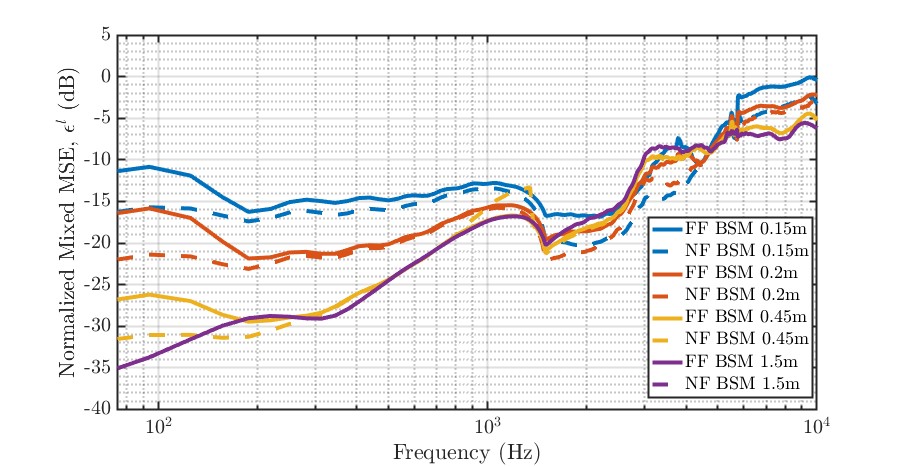}
   \caption{Normalized Mixed MSE of the left ear for far-field (solid lines) and -field (dashed lines) BSM filters as a function of frequency for source distances ranging from 0.15 to 1.5 meters, without head rotations.}
    \label{fig:Mixed_MSE}
\end{figure}

Figure~\ref{fig:Mixed_MSE} presents the normalized Mixed MSE of the left ear, as defined in Eq.~(\ref{mixed_error}), for both far-field and near-field BSM filters, computed as detailed in sec. \ref{NF BSM}. Since the results are identical for both ears in this symmetric scenario, only left ear is shown. The near-field results are indicated by dashed lines, while the far-field results are shown using solid lines. The analysis covers frequencies from \(75 \, \text{Hz}\) to \(10\, \text{kHz}\) for several source distances. The distance \(r_s = 1.5 \, \text{m}\) corresponds to the far-field distance in this simulation.

Across most of the frequency range and source distances, the far-field BSM—which assumes a fixed source distance of 1.5~m—yields slightly higher normalized MSE values than the near-field BSM. This suggests that under ideal listening conditions (i.e., without head rotation), incorporating distance-dependent information via the near-field model provides a marginal improvement in reproduction accuracy. For very close sources (\(r_s = 0.15 \, \text{m}\) and \(r_s = 0.2 \, \text{m}\)), approximately 5--10~cm from the surface of the head, the error increases significantly for both filters, highlighting the limitations of BSM in extreme near-field conditions.

\begin{figure}[t]
    \centering
    \includegraphics[trim={45pt 0 0 0},clip,width=0.5\textwidth]{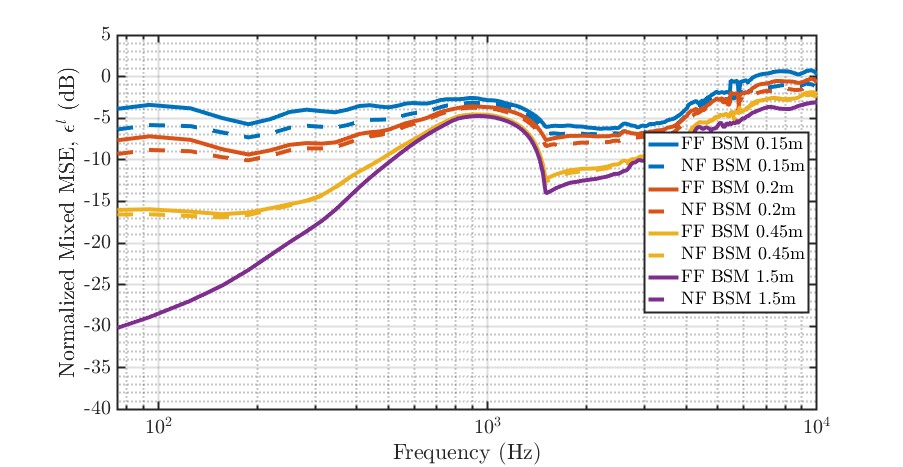}\\[-0.5ex]
    \includegraphics[trim={45pt 0 0 0},clip,width=0.5\textwidth]{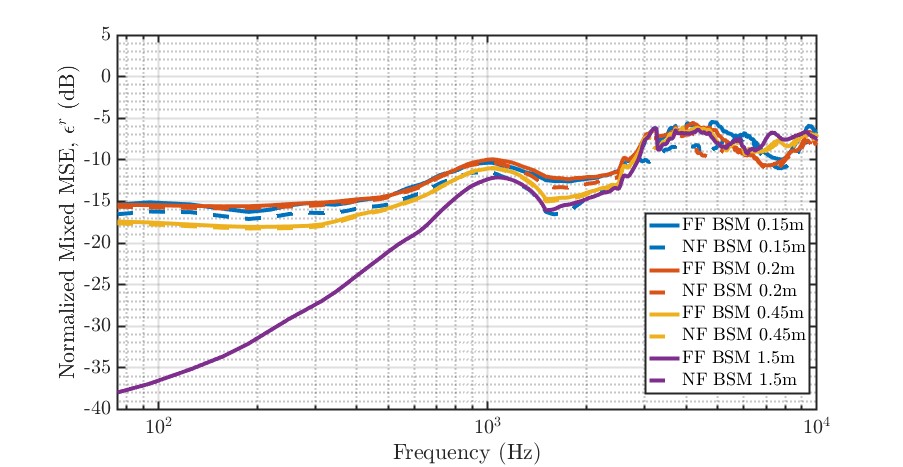}
    \caption{Same as figure 2 but under head rotation of \(40^\circ\) for the left ear (top) and right ear (bottom).}
    \label{fig:rotations_40}
\end{figure}

Figure~\ref{fig:rotations_40} shows the normalized Mixed MSE under azimuthal head rotation of \(40^\circ\). The asymmetry introduced by rotation is evident in the differences between the left and right ears. Despite this, the performance gap between the near-field and far-field BSM remains relatively small under head rotation, suggesting that the added value of incorporating source-distance information is limited in dynamic scenarios as well.

Overall, the results indicate that while the near-field BSM consistently outperforms its far-field counterpart, the improvement is modest under most conditions. Both filters exhibit degraded performance at very short distances, underscoring the challenge of achieving accurate binaural reproduction for extremely close sources. This aspect will be investigated in more detail in Section~\ref{Null Space}.

%\begin{figure}[h!]
%    \hspace{-1 cm}
% \includegraphics[trim={50pt 0 0 0},clip,width=0.6\textwidth]{PS_weights_no_head_rot.jpg}
%    \caption{Normalized MSE for Near-Field BSM filters for source distances ranging from 0.15 to 3.2 meters.}
%    \label{fig:Pw_weights_no_head_rot}
%\end{figure}

\subsubsection{ILD error}
\begin{figure}[t]
    \centering
 \setlength{\abovecaptionskip}{5pt} % Space above caption
    \setlength{\belowcaptionskip}{-5pt} % Space below caption
    \setlength{\intextsep}{5pt} % Space above/below in-text figures

    \begin{subfigure}[t]{0.4\textwidth}
        \includegraphics[width=\textwidth]{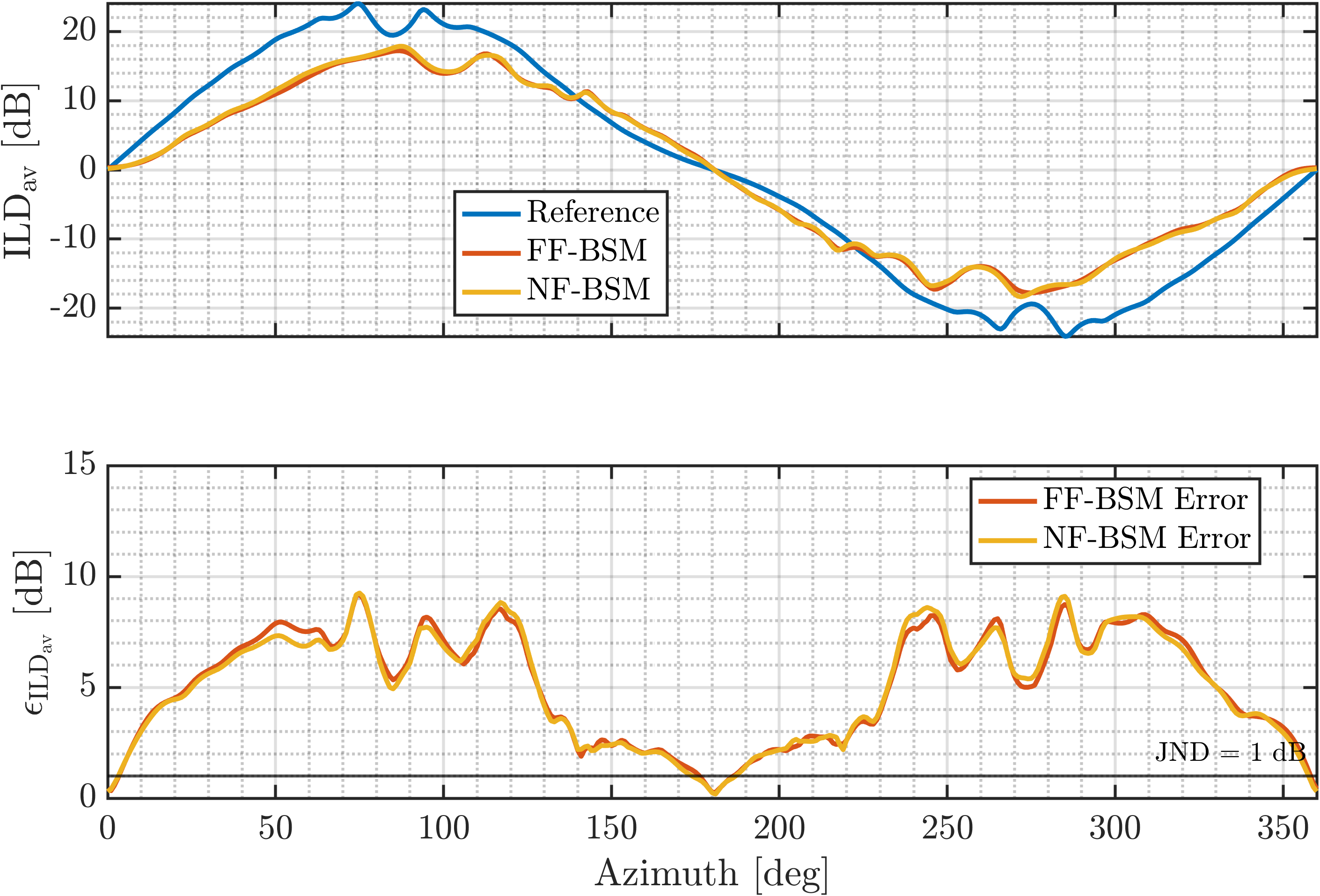}
        \caption{\(r_s = 0.45\, \text{m}\)}
    \end{subfigure}
    \iffalse
    \vspace{1.5em}

    \begin{subfigure}[t]{0.4\textwidth}
        \includegraphics[width=\textwidth]{figs/ILD/ILD_SIMATF_SimHRTF_Compare_0.20.png}
        \caption{\(r_s = 0.2\, \text{m}\)}
    \end{subfigure}

    \vspace{1.5em}
\fi
    \begin{subfigure}[t]{0.4\textwidth}
        \includegraphics[width=\textwidth]{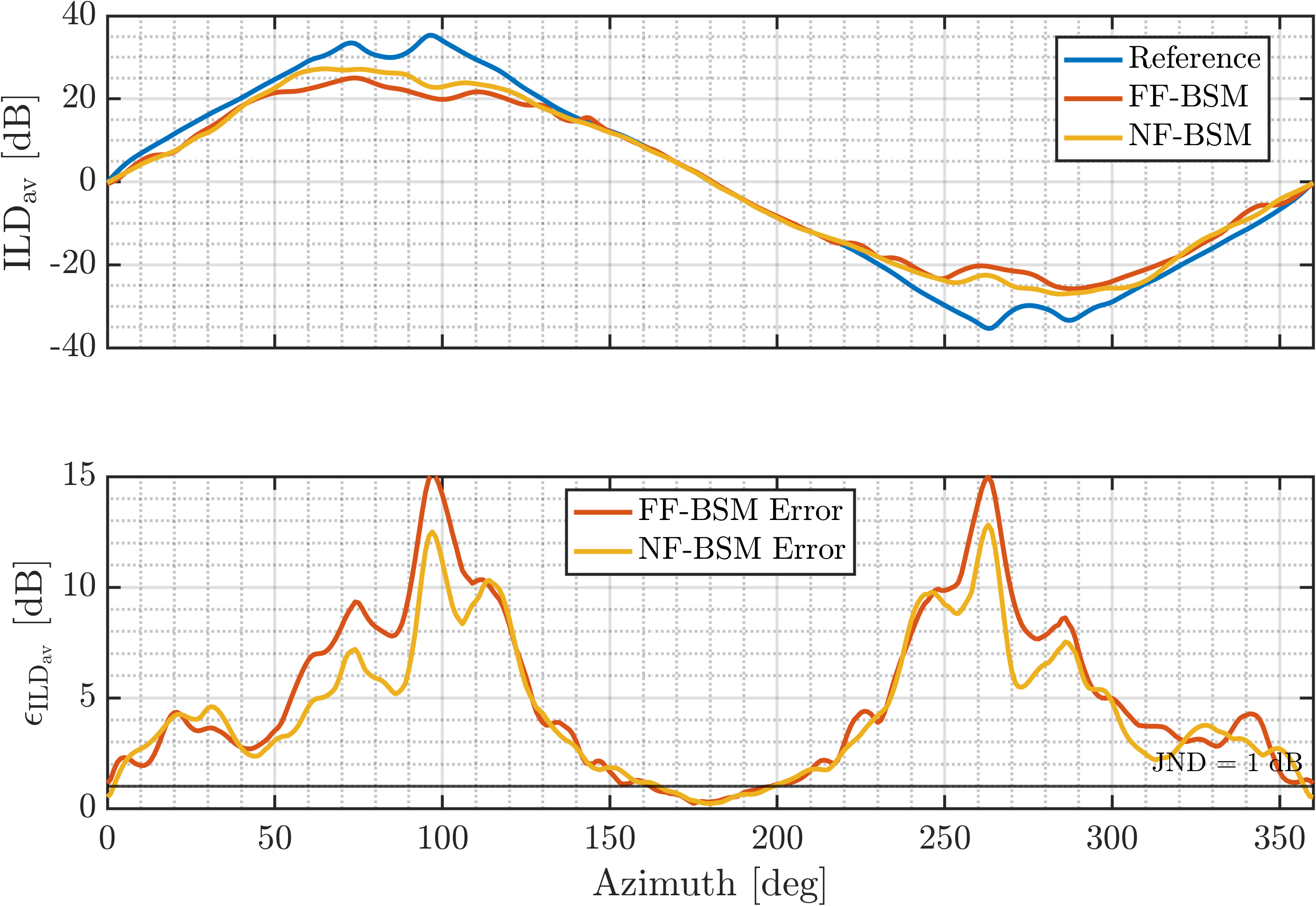}
        \caption{\(r_s = 0.15\, \text{m}\)}
    \end{subfigure}

    \caption{ILD values and ILD errors for far-field and near-field BSM filters at source distances of 0.15\,m and 0.45\,m.}
    \label{fig:ILD_comparison}
\end{figure}

The ILD values and ILD errors were calculated according to Eqs.~(\ref{ILD1})–(\ref{ILD3}). The results for both far-field and near-field BSM are shown in Fig.~\ref{fig:ILD_comparison} for source distances of 0.15 m and 0.45 m. Small differences are observed between the two methods, suggesting that both far-field and near-field BSM achieve comparable performance in preserving ILD. This consistency indicates that incorporating either far-field or near-field BSM leads to similar accuracy in ILD. Note that both errors are significantly above the just noticeable difference (JND) threshold of 1 dB \cite{blauert1997spatial,yost1988discrimination}, suggesting that the errors are clearly audible.

\captionsetup[subfigure]{justification=centering, skip=6pt} % add vertical space between plot and caption

\subsubsection{ITD error}

\begin{figure}[t]
    \centering
    \setlength{\abovecaptionskip}{5pt} % Space above caption
    \setlength{\belowcaptionskip}{-5pt} % Space below caption
    \setlength{\intextsep}{5pt} % Space above/below in-text figures

    \begin{subfigure}[t]{0.4\textwidth}
        \includegraphics[width=\textwidth]{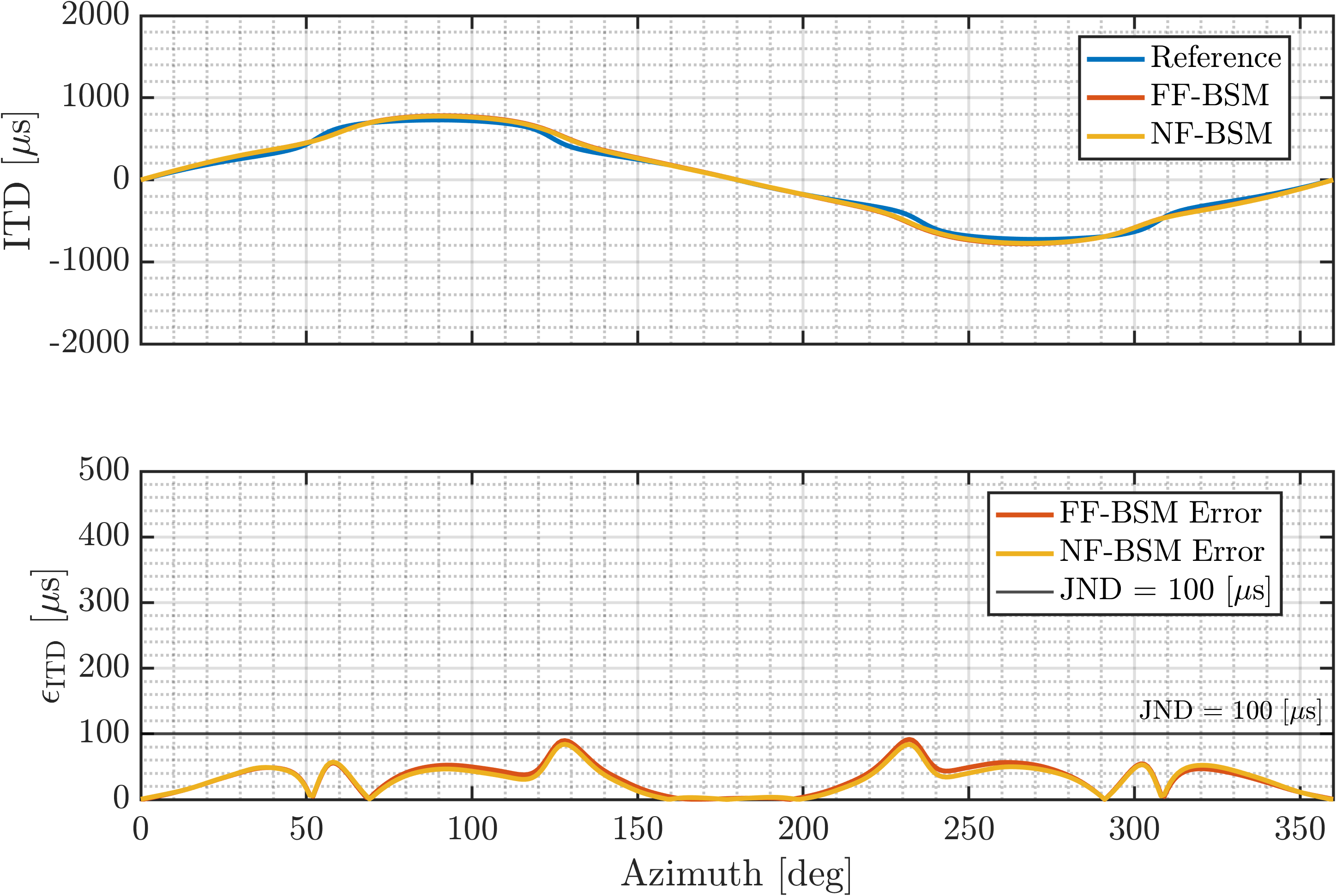}
        \caption{\(r_s = 0.45\, \text{m}\)}
    \end{subfigure}

    \vspace{1.5em}
    \iffalse
    \begin{subfigure}[t]{0.4\textwidth}
        \includegraphics[width=\textwidth]{figs/ITD/ITD_SIMATF_SimHRTF_Compare_0.20.png}
        \caption{\(r_s = 0.2\, \text{m}\)}
    \end{subfigure}

    \vspace{1.5em}
    \fi
    \begin{subfigure}[t]{0.4\textwidth}
        \includegraphics[width=\textwidth]{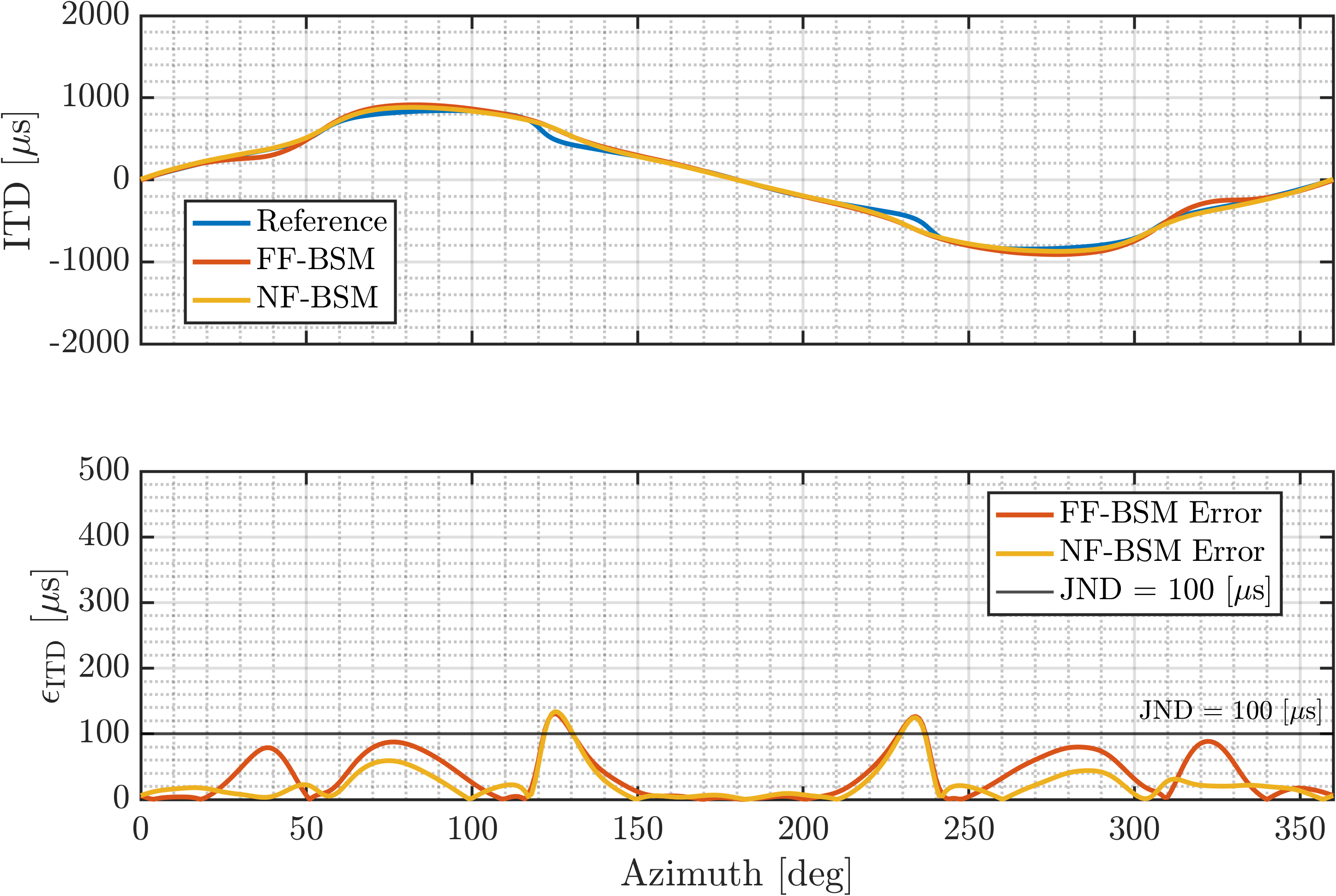}
        \caption{\(r_s = 0.15\, \text{m}\)}
    \end{subfigure}

    \caption{ITD values and ITD errors for far-field and near-field BSM filters at source distances of 0.15\,m and 0.45\,m.}
    \label{fig:ITD_comparison}
\end{figure}

The ITD was calculated following the method described in Eqs.~(\ref{Tau}) and (\ref{itd_est}). 
Fig.~\ref{fig:ITD_comparison} shows the frequency-averaged ITD of the reference and BSM signals. 
The corresponding frequency-averaged ITD error, defined in Eq.~(\ref{ITDcalcc}), is also displayed as a function of azimuth and computed using both far-field and near-field BSM filters for the two tested source distances (0.15\,m and 0.45\,m). 

The ITD errors are relatively small for both far-field and near-field BSM, remaining below the JND threshold of $20 \,\mu\text{s}$ for frontal directions and $100 \,\mu\text{s}$ for lateral directions~\cite{mossop1998lateralization,andreopoulou2017identification}.

\vspace*{1.5em}  % Align with ITD figure

\captionsetup[subfigure]{justification=centering, skip=6pt} % add vertical space between plot and caption

\subsubsection{Null Space projection}
\label{Null Space}
As observed in the previous section, source distances of only a few centimeters from the microphone array result in significant binaural errors for both far-field and near-field BSM. Even when the source distance is known and near-field BSM is applied, substantial performance degradation persists. This indicates that incorporating source-distance information alone is insufficient to fully resolve the inaccuracies in extreme near-field conditions.

 As described in Eq. (\ref{weights}), BSM aims to find a combination of microphone steering vectors that reconstructs the HRTF. When the HRTF lies within the steering matrix range, the BSM filters can achieve an accurate reconstruction, and a small error. 

This condition can be quantified using the \textit{Null Space Projection Measure} defined in Eq. (\ref{null_measure}). This measure evaluates how much of the HRTF vector lies within the \textit{null space} of the steering matrix—essentially, the portion of the HRTF that cannot be represented using the given microphone array. 

\begin{figure}[!htbp]
    \centering
    \includegraphics[trim={45pt 0 0 0},clip,width=0.5\textwidth]{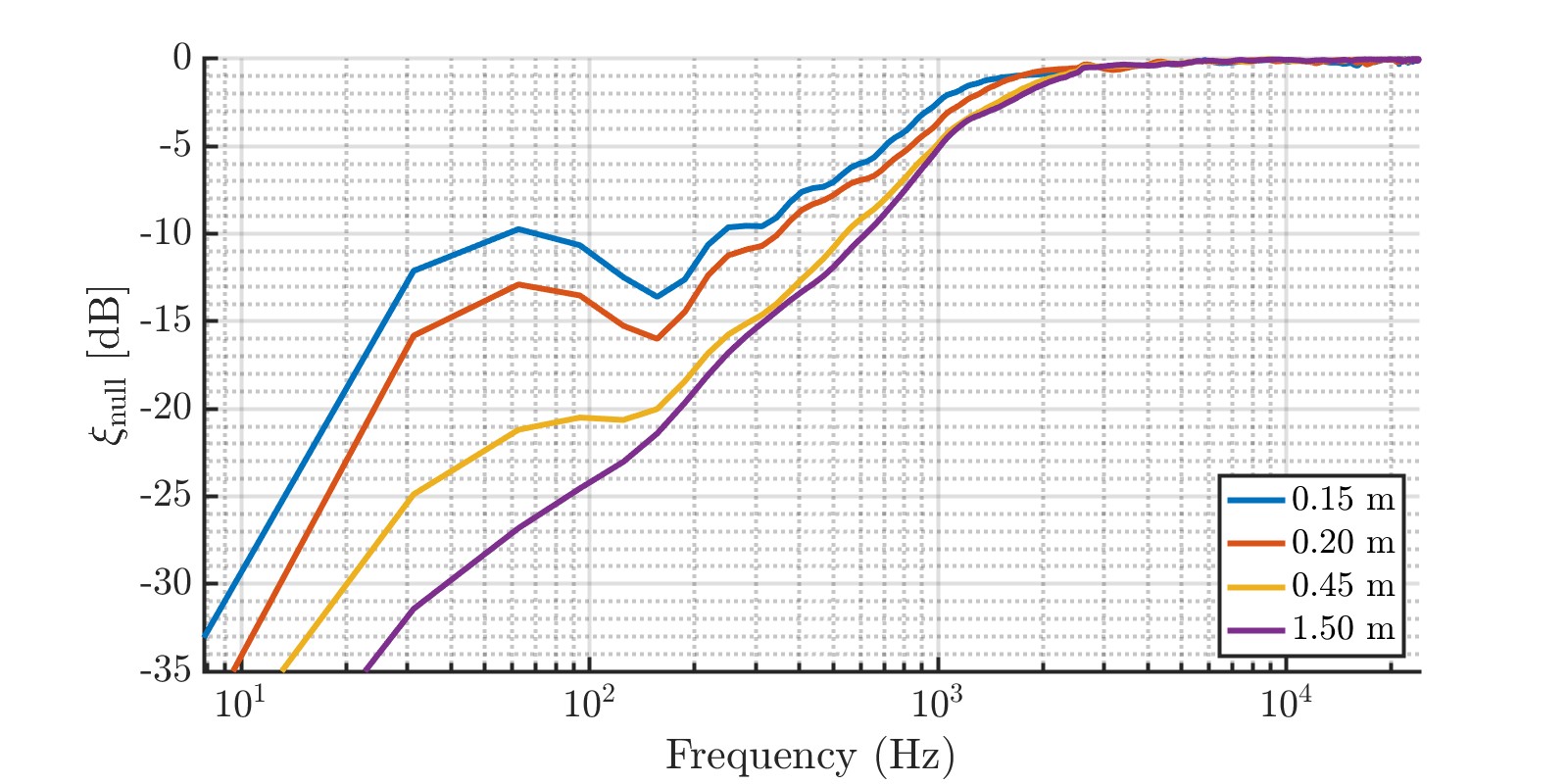}
    \caption{Null space projection values, as in Eq.~(\ref{null_measure}), of the near-field HRTF onto the near-field steering matrix, evaluated across different source distances, where the distance of 1.5\,m corresponds to the far-field case.
}
    \label{fig:null_projection}
\end{figure}

Fig. \ref{fig:null_projection} presents the null space projection values of the near-field HRTF onto the near-field steering matrix across different source distances, along with the projection of the far-field HRTF onto the far-field steering matrix. The results reveal that for sources located at very short distances, the null space projection is significantly larger, indicating that a considerable portion of the HRTF is not representable by the steering matrix. The underlying reason for this dependency on source distance is not yet fully understood. Further investigation into this phenomenon is left for future research.

\subsection{Results: BSM with limited FoV}
 \label{FoV}
In Chapter~\ref{FOV Design}, we introduced the FoV design approach, which incorporates a directional prior based on the assumption that sources of interest are located within a frontal region. This design introduces a tunable parameter \(\beta\) that controls the trade-off between accurate reproduction inside and outside the FoV. In the following results, we present the performance for \(\beta = 0.2\), which was found to offer the best trade-off between reducing errors within the FoV while avoiding excessive degradation outside it. A FoV of $\pm 45^\circ$ was selected in both azimuth and elevation around the front looking direction.

 \subsubsection{Binaural Error with FoV weighting}

\begin{figure}[t]
    \centering
    \begin{subfigure}[b]{0.99\linewidth}
        \centering
        \includegraphics[width=\textwidth]{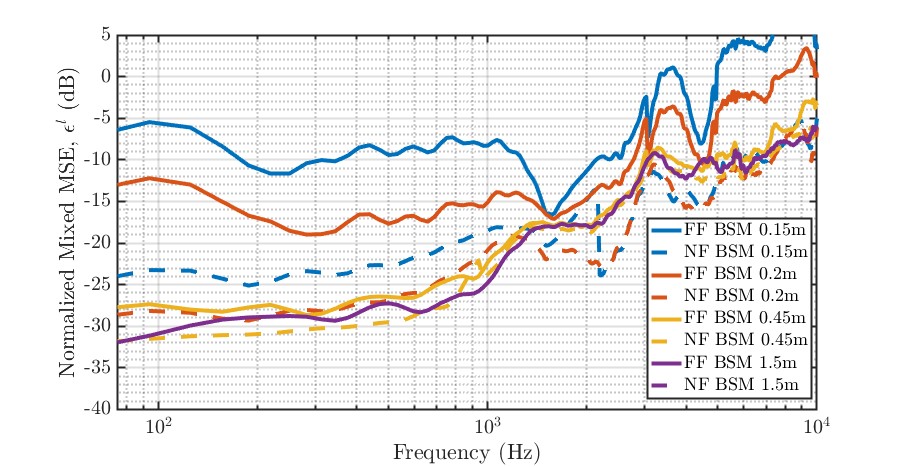}
        \caption{Directions within FoV (\(\pm 45^\circ\) azimuth and elevation)}
        \label{fig:inside_fov}
    \end{subfigure}

    \vspace{2em}

    \begin{subfigure}[b]{0.99\linewidth}
        \centering
        \includegraphics[width=\textwidth]{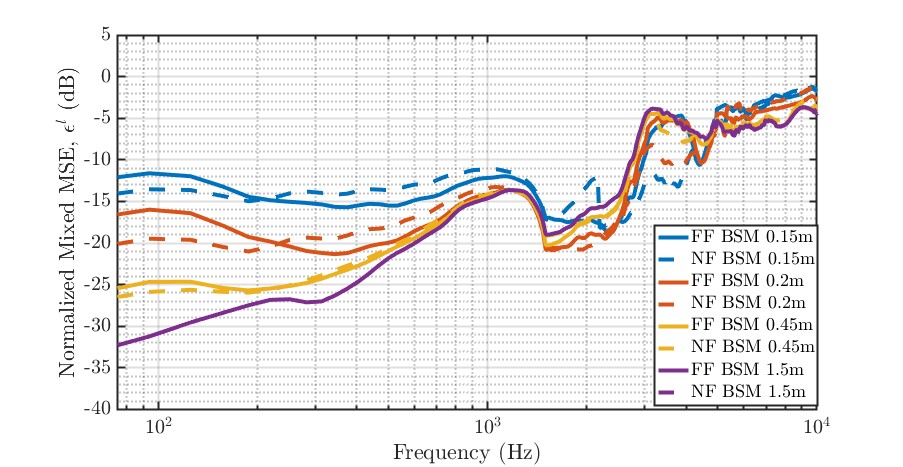}
        \caption{Directions outside FoV}
        \label{fig:outside_fov}
    \end{subfigure}

    \caption{Normalized Mixed MSE of the left ear for far-field and near-field BSM filters under Field of View weighting, (a) inside the FoV, and (b) outside the FoV.}
    \label{fig:fov_binaural_error}
\end{figure}

Fig. ~\ref{fig:fov_binaural_error} presents the normalized Mixed MSE for the left ear, comparing the performance of far-field and near-field BSM filters under FoV weighting. The results are shown separately for DOAs inside the FoV and outside the FoV, in two subfigures.

In the FoV region (top figure), the differences between the near-field and far-field BSM are significantly more pronounced compared to the unweighted (spatially uniform) case shown in Fig. ~\ref{fig:Mixed_MSE}. The near-field BSM yields notably lower binaural errors, especially below approximately 2~kHz, while the far-field BSM performance degrades sharply in this frequency range. This result highlights the importance of incorporating source-distance information when source directions are focused within a limited region.

In contrast, the results for sources outside the FoV (bottom figure) remain largely similar to those in the unweighted case. The performance gap between the far-field and near-field BSM is relatively small, and both filters show comparable trends to those in Fig. ~\ref{fig:Mixed_MSE}. This consistency suggests that the directional prior primarily enhances performance within the attended region but does not significantly distort the reproduction quality outside it.

\begin{figure*}[t]
    \centering
    \begin{subfigure}[b]{0.48\textwidth}
        \centering
        \includegraphics[width=\textwidth]{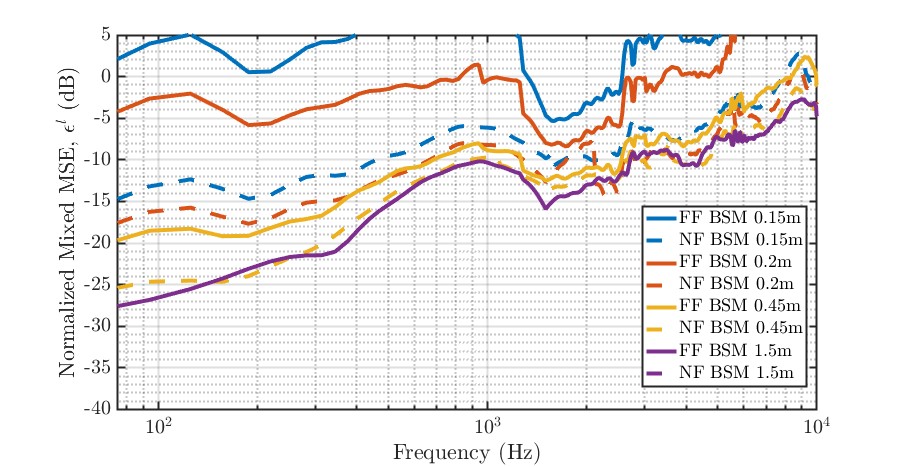}
        \caption{Directions within FoV -- Left Ear}
    \end{subfigure}
    \hfill
    \begin{subfigure}[b]{0.48\textwidth}
        \centering
        \includegraphics[width=\textwidth]{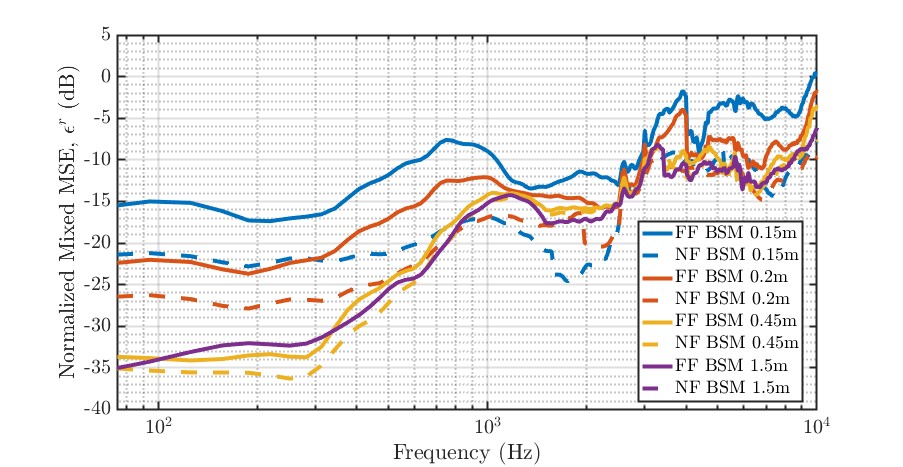}
        \caption{Directions within FoV -- Right Ear}
    \end{subfigure}
    
    \vspace{1em}
    
    \begin{subfigure}[b]{0.48\textwidth}
        \centering
        \includegraphics[width=\textwidth]{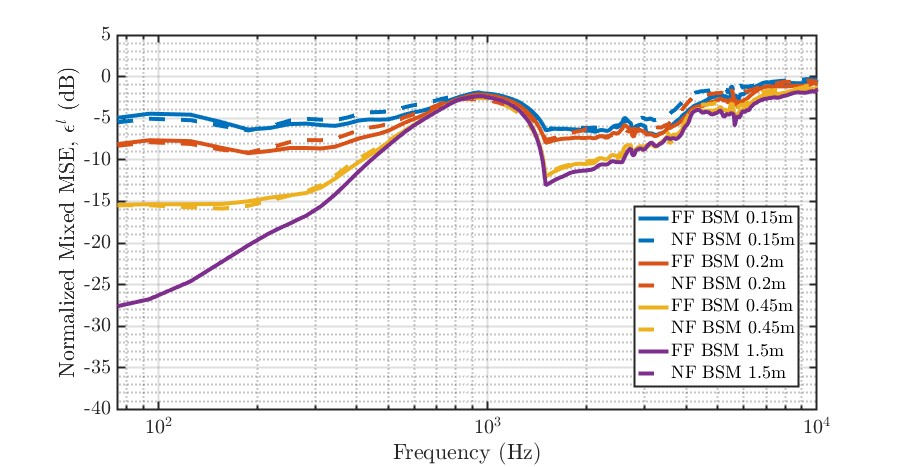}
        \caption{Directions outside FoV -- Left Ear}
    \end{subfigure}
    \hfill
    \begin{subfigure}[b]{0.48\textwidth}
        \centering
        \includegraphics[width=\textwidth]{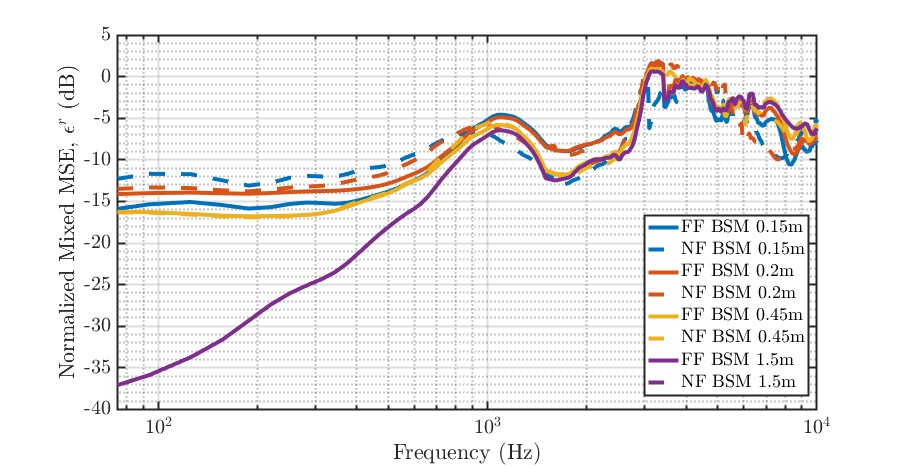}
        \caption{Directions outside FoV - Right Ear}
    \end{subfigure}

    \caption{Normalized Mixed MSE for far-field and near-field BSM filters under Field of View weighting and a \(40^\circ\) head rotation, (a) inside the FoV -- left ear, (b) inside the FoV -- right ear, (c) outside the FoV -- left ear, and (d) outside the FoV -- right ear.
}
    \label{fig:fov_head_rotation}
\end{figure*}

Fig. ~\ref{fig:fov_head_rotation} shows the Mixed error results under a head rotation of \(40^\circ\). Since the head is no longer front-facing, the left–right symmetry is broken, and results are presented for both ears. In the FoV region (top figures), the difference between the far-field and near-field BSM filters becomes more pronounced when head rotation is introduced, particularly for the ear oriented away from the microphone array. The near-field BSM continues to yield superior performance, especially in the low-frequency range, while the far-field filter shows notable degradation.

Outside the FoV (bottom figures), the results remain largely consistent with those observed in the spatially uniform case under \(40^\circ\) head rotation (see Fig. ~\ref{fig:rotations_40}), suggesting that the directional weighting has minimal impact on performance for off-FoV sources.

\subsubsection{ILD Error with FoV weighting}

\begin{figure}[t]
    \centering
    \setlength{\abovecaptionskip}{5pt} % Space above caption
    \setlength{\belowcaptionskip}{-5pt} % Space below caption
    \setlength{\intextsep}{5pt} % Space above/below in-text figures

    \begin{subfigure}[t]{0.4\textwidth}
        \includegraphics[width=\textwidth]{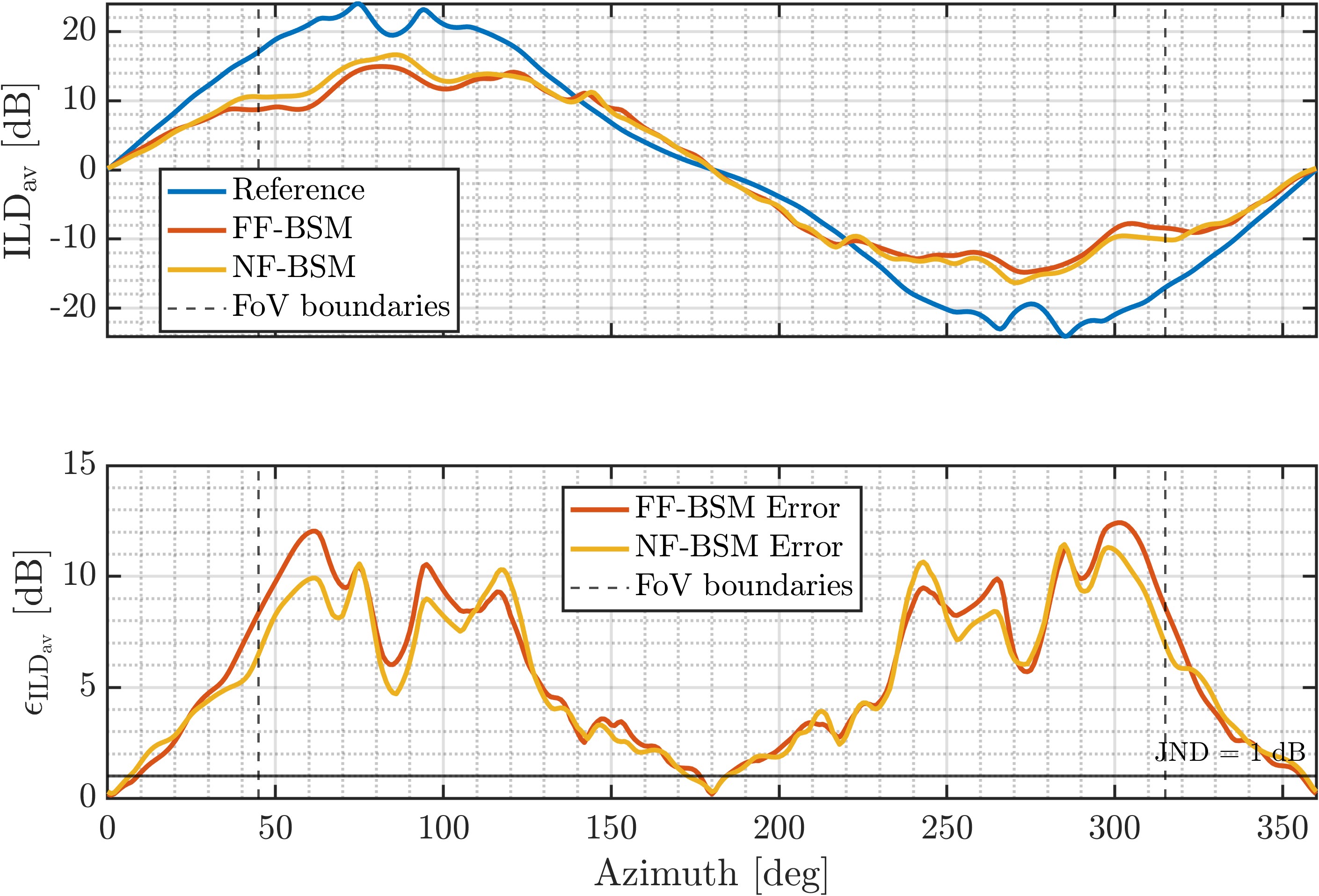}
        \caption{\(r_s = 0.45\,\text{m}\)}
        \label{fig:ILD_FOV_045}
    \end{subfigure}

    \vspace{1.5em}
\iffalse
    \begin{subfigure}[t]{0.4\textwidth}
        \includegraphics[width=\textwidth]{figs/FOV/ILD/ILD_SIMATF_SimHRTF_Compare_FOV_0.20_rot0.0_FoVbound.jpg}
        \caption{\(r_s = 0.20\,\text{m}\)}
        \label{fig:ILD_FOV_020}
    \end{subfigure}

    \vspace{1.5em}
\fi
    \begin{subfigure}[t]{0.4\textwidth}
        \includegraphics[width=\textwidth]{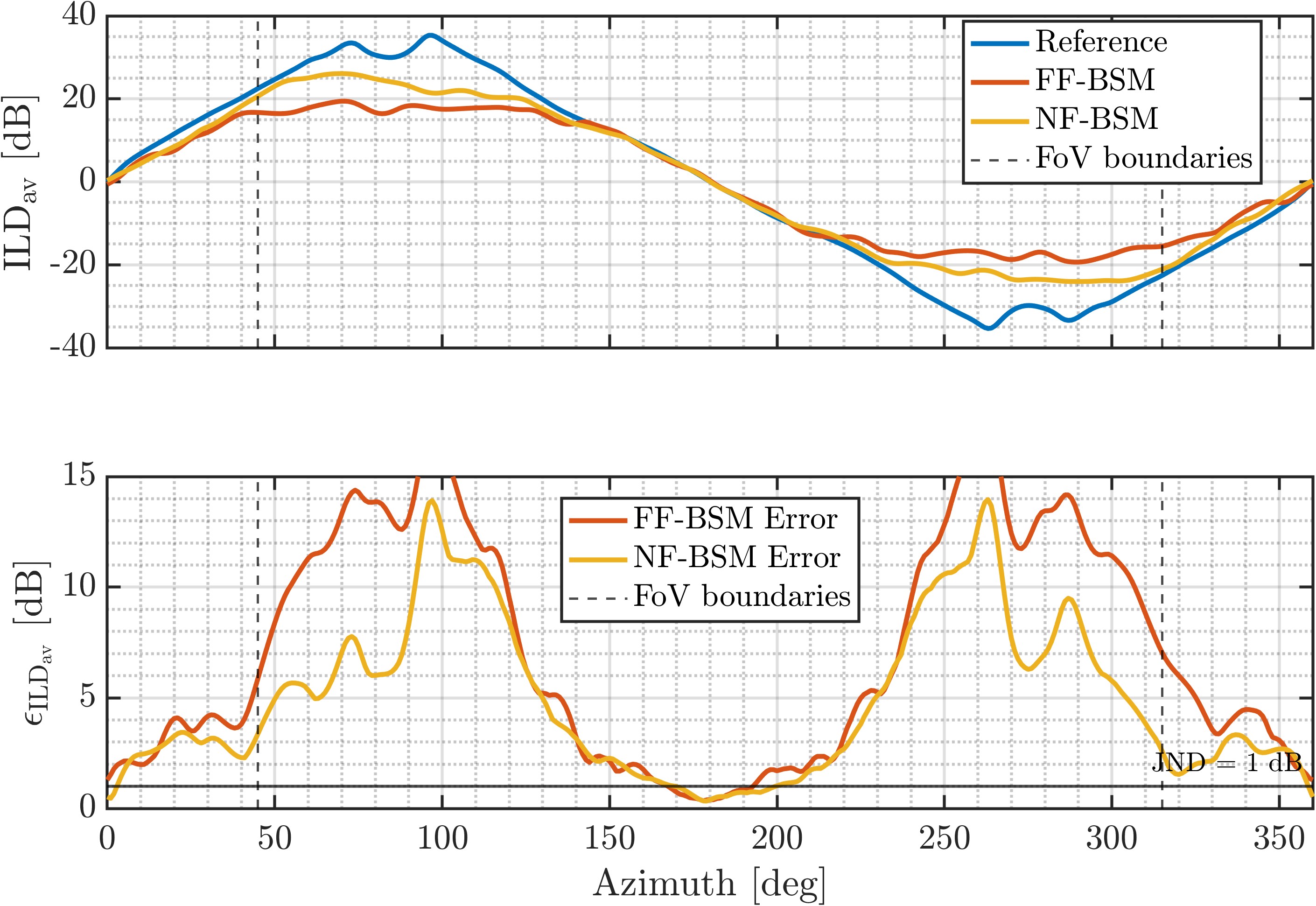}
        \caption{\(r_s = 0.15\,\text{m}\)}
        \label{fig:ILD_FOV_015}
    \end{subfigure}

    \caption{Average ILD values and ILD errors for far-field and near-field BSM filters under Field of View weighting, shown for source distances of 0.15\,m and 0.45\,m. The dashed vertical lines in the figure indicate the boundaries of the FoV region (\(\pm 45^\circ\) from the center, i.e., at \(45^\circ\) and \(315^\circ\)).
}
    \label{fig:ILD_FOV}
\end{figure}

Fig.~\ref{fig:ILD_FOV} shows the ILD values and ILD errors for near-field and far-field BSM filters at source distances of 0.15\,m and 0.45\,m without head rotation with FoV weighting. The dashed vertical lines in the figure indicate the boundaries of the FoV region.
 The near-field BSM consistently achieves better results, with lower ILD errors overall. Inside the FoV, the error is significantly smaller, and similarly, reduced errors are observed at the back of the head, possibly due to the symmetry between the front and back regions in this scenario. In contrast, outside the FoV—particularly at the side directions—the ILD errors are higher with both filters. The differences between near-field and far-field BSM are more pronounced compared to the spatially uniform case, shown in Fig.~\ref{fig:ILD_comparison} especially for close source distances.

\begin{figure}[t]
    \centering
    \setlength{\abovecaptionskip}{5pt} % Space above caption
    \setlength{\belowcaptionskip}{-5pt} % Space below caption
    \setlength{\intextsep}{5pt} % Space above/below in-text figures

    \begin{subfigure}[t]{0.4\textwidth}
        \includegraphics[width=\textwidth]{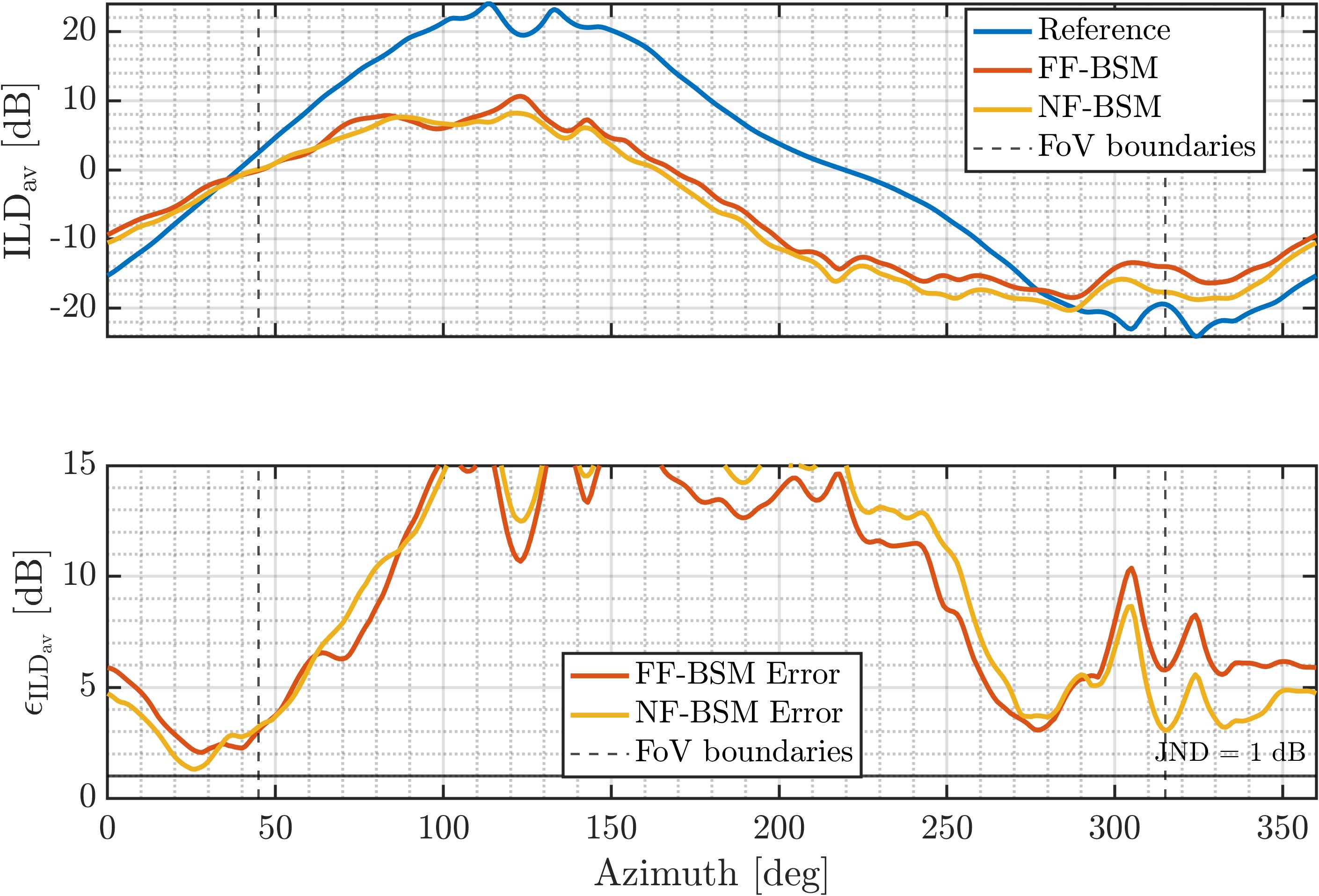}
        \caption{\(r_s = 0.45\,\text{m}\)}
        \label{fig:ILD_FOV_rot_045}
    \end{subfigure}

    \vspace{1.5em}
\iffalse
    \begin{subfigure}[t]{0.4\textwidth}
        \includegraphics[width=\textwidth]{figs/FOV/ILD/ILD_SIMATF_SimHRTF_Compare_FOV_0.20_rot0.7_FoVbound.jpg}
        \caption{\(r_s = 0.20\,\text{m}\)}
        \label{fig:ILD_FOV_rot_020}
    \end{subfigure}

    \vspace{1.5em}
\fi
    \begin{subfigure}[t]{0.4\textwidth}
        \includegraphics[width=\textwidth]{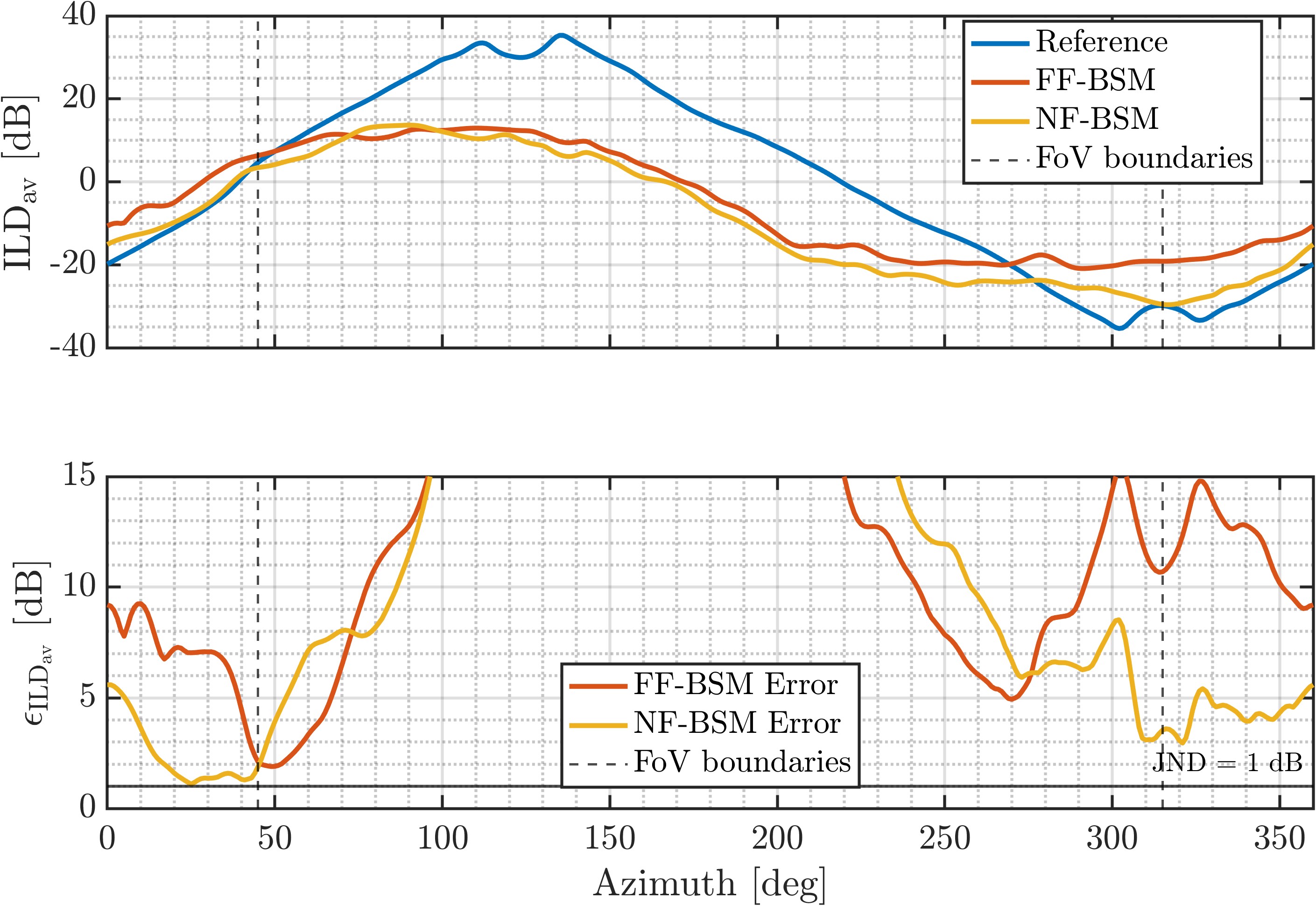}
        \caption{\(r_s = 0.15\,\text{m}\)}
        \label{fig:ILD_FOV_rot_015}
    \end{subfigure}

    \caption{Same as figure 9 but under head rotation of \(40^\circ\).}
    \label{fig:ILD_FOV_rot}
\end{figure}

Fig.~\ref{fig:ILD_FOV_rot} presents the ILD and ILD error results under a \(40^\circ\) head rotation. As seen, there is a clear distinction between the inside-FoV and outside-FoV regions. Within the FoV, the differences between near-field and far-field BSM become even larger, particularly for the very close source distance. Additionally, the errors are generally larger compared to the case without head rotation, highlighting the increased challenges introduced by head movement.

\captionsetup[subfigure]{justification=centering, skip=6pt} % add vertical space between plot and caption
\captionsetup[subfigure]{justification=centering, skip=6pt} % add vertical space between plot and caption

\subsubsection{ITD Error with FoV weighting}

\begin{figure}[t]
    \centering
    \setlength{\abovecaptionskip}{5pt}
    \setlength{\belowcaptionskip}{-5pt}
    \setlength{\intextsep}{5pt}

    \begin{subfigure}[t]{0.4\textwidth}
        \includegraphics[width=\textwidth]{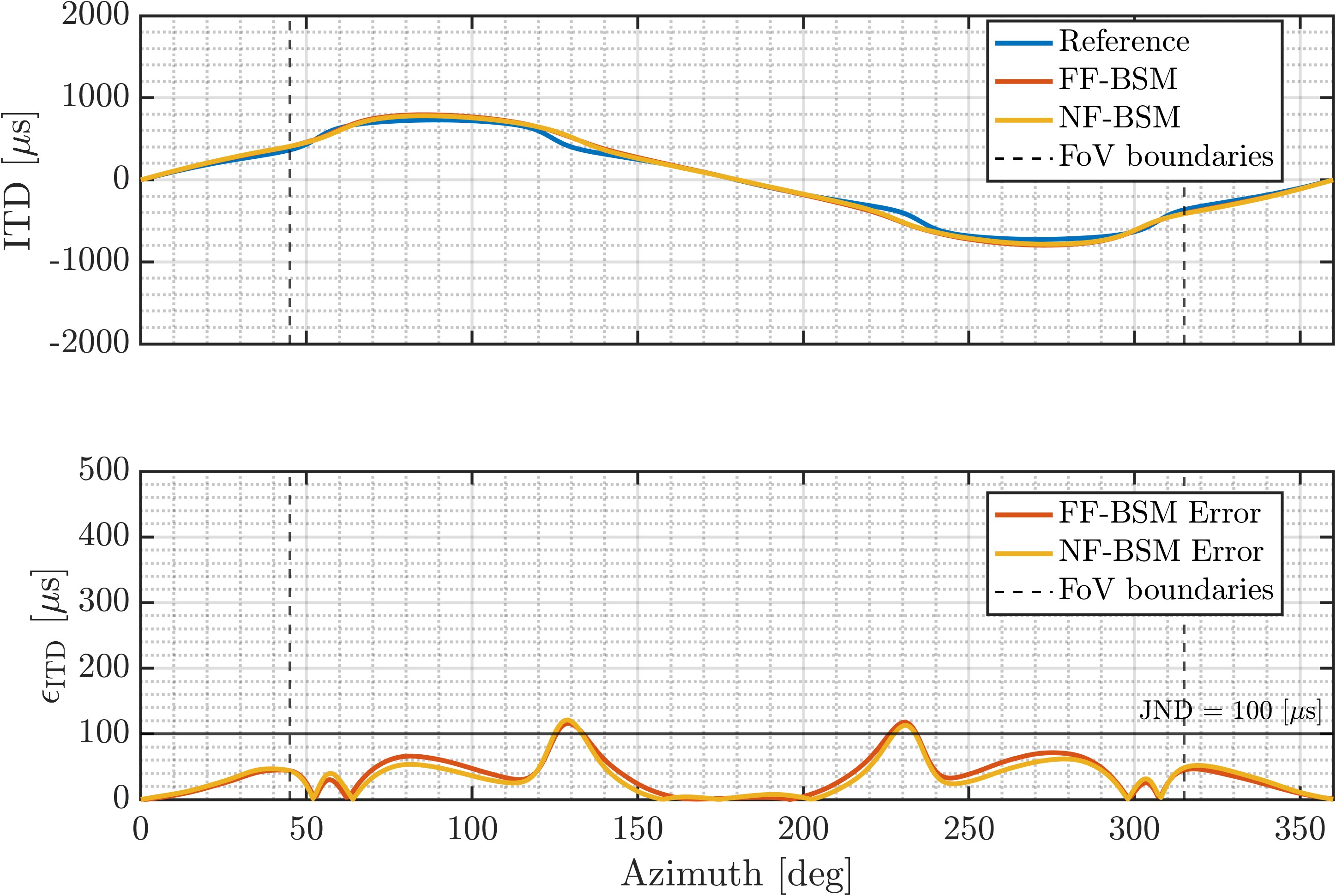}
        \caption{\(r_s = 0.45\,\text{m}\)}
        \label{fig:ITD_FOV_045}
    \end{subfigure}

    \vspace{1.5em}
\iffalse
    \begin{subfigure}[t]{0.4\textwidth}
        \includegraphics[width=\textwidth]{figs/FOV/ITD/ITD_SIMATF_SimHRTF_Compare_FOV0.20_rot0.0_FoVbound.jpg}
        \caption{\(r_s = 0.20\,\text{m}\)}
        \label{fig:ITD_FOV_020}
    \end{subfigure}

    \vspace{1.5em}
\fi
    \begin{subfigure}[t]{0.4\textwidth}
        \includegraphics[width=\textwidth]{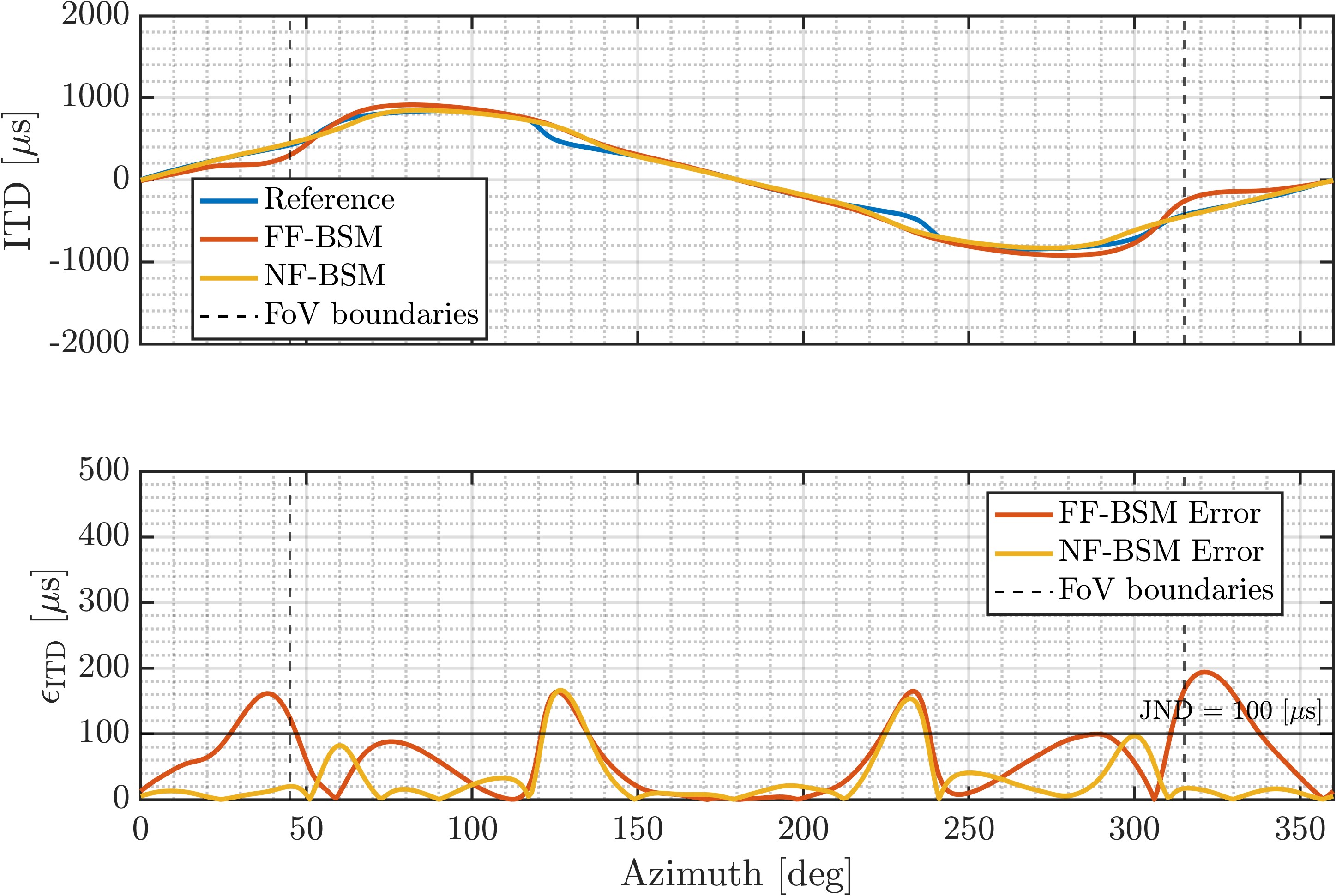}
        \caption{\(r_s = 0.15\,\text{m}\)}
        \label{fig:ITD_FOV_015}
    \end{subfigure}

    \caption{Average ITD values and ITD errors for far-field and near-field BSM filters under Field of View weighting without head rotation, shown for source distances of 0.15\,m and 0.45\,m. The dashed vertical lines in the figure indicate the boundaries of the FoV region (\(\pm 45^\circ\) from the center, i.e., at \(45^\circ\) and \(315^\circ\)).
}
    \label{fig:ITD_FOV}
\end{figure}

Fig.~\ref{fig:ITD_FOV} shows the ITD values and ITD errors for near-field and far-field BSM filters under FoV weighting without head rotation. The dashed vertical lines in the figure indicate the boundaries of the FoV region. The errors achieved with the near-field BSM are generally quite small—usually below the JND line of 100~$\mu$s inside the FoV, and slightly above it outside the FoV. The differences between near-field and far-field BSM are typically not very dominant in this configuration.

\begin{figure}[t]
    \centering
    \setlength{\abovecaptionskip}{5pt}
    \setlength{\belowcaptionskip}{-5pt}
    \setlength{\intextsep}{5pt}

    \begin{subfigure}[t]{0.4\textwidth}
        \includegraphics[width=\textwidth]{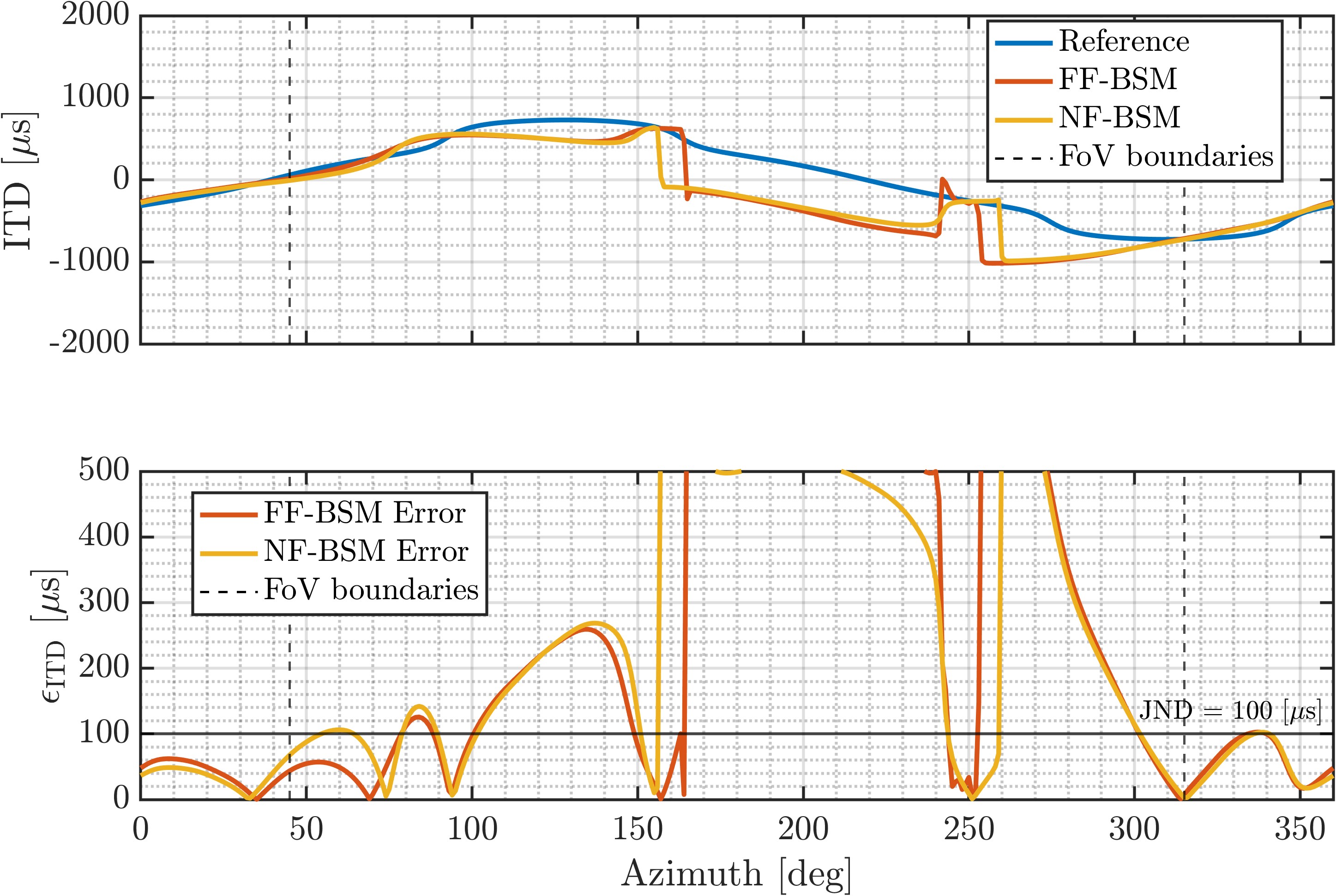}
        \caption{\(r_s = 0.45\,\text{m}\)}
        \label{fig:ITD_FOV_rot_045}
    \end{subfigure}

    \vspace{1.5em}
\iffalse
    \begin{subfigure}[t]{0.4\textwidth}
        \includegraphics[width=\textwidth]{figs/FOV/ITD/ITD_SIMATF_SimHRTF_Compare_FOV0.20_rot0.7_FoVbound.jpg}
        \caption{\(r_s = 0.20\,\text{m}\)}
        \label{fig:ITD_FOV_rot_020}
    \end{subfigure}

    \vspace{1.5em}
\fi
    \begin{subfigure}[t]{0.4\textwidth}
        \includegraphics[width=\textwidth]{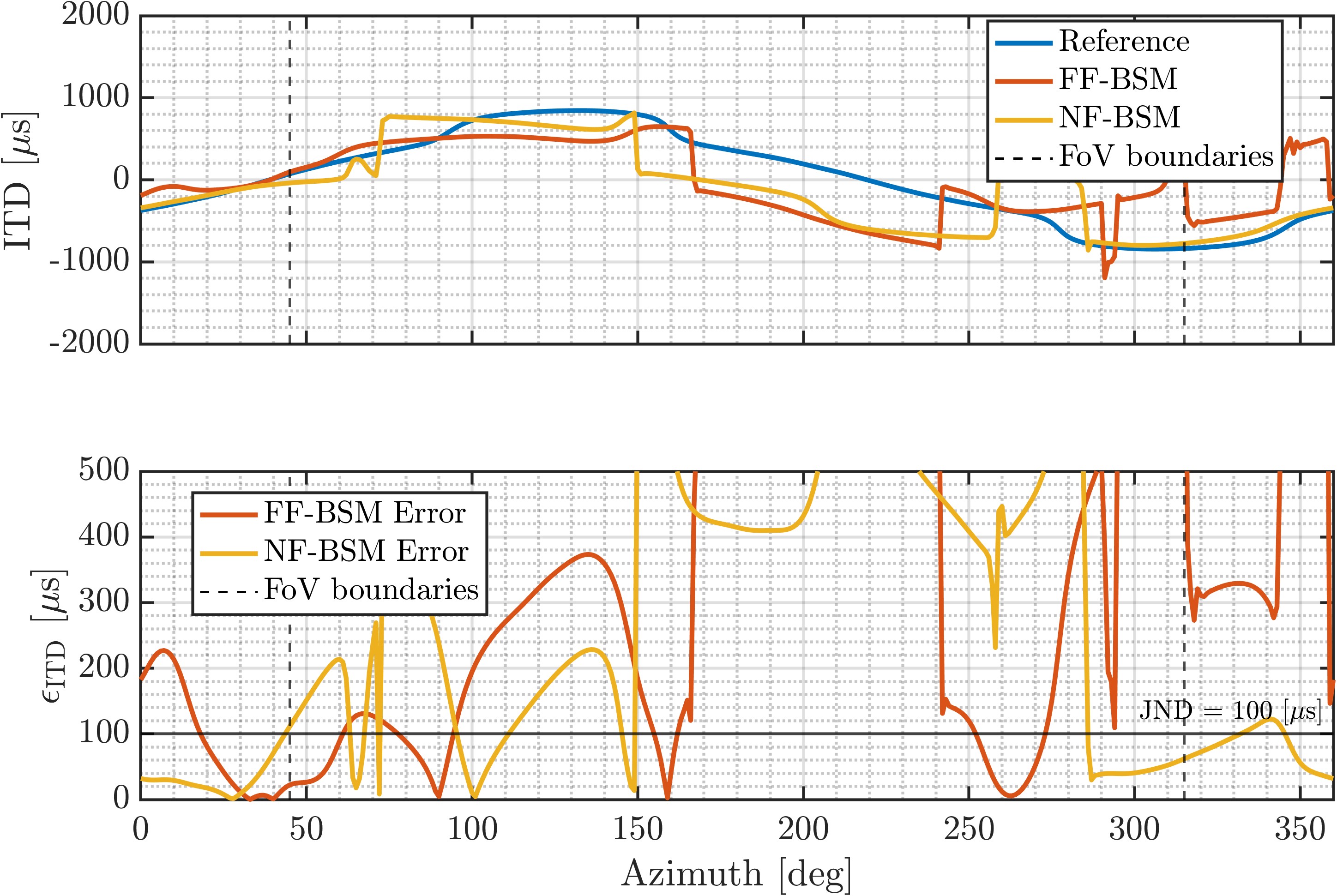}
        \caption{\(r_s = 0.15\,\text{m}\)}
        \label{fig:ITD_FOV_rot_015}
    \end{subfigure}

    \caption{Same as figure 11 but under head rotation of \(40^\circ\) for the left ear (top) and right ear (bottom).}
    \label{fig:ITD_FOV_rot}
\end{figure}

However, Fig. ~\ref{fig:ITD_FOV_rot} shows the corresponding results under a head rotation of \(40^\circ\). In this case, the ITD errors are noticeably larger than in the static head condition across all scenarios. In this case, the near-field BSM  yields noticeably smaller errors, particularly in the directions inside the FoV, emphasizing its robustness and accuracy even under head movement.

\captionsetup[subfigure]{justification=centering, skip=6pt}

\captionsetup[subfigure]{justification=centering, skip=6pt}

\iffalse
\subsubsection{Null Space Projection within the FoV}

Fig.~\ref{fig:null_projection_fov} shows the null space projection values when the analysis is restricted strictly to directions within the FoV. Specifically, both the HRTF vector and the steering matrix are truncated to include only the directions that lie within the FoV, effectively setting the contributions from directions outside the FoV to zero in this projection computation.

Compared to the original projection results shown in Fig.~\ref{fig:null_projection}, the values in Fig.~\ref{fig:null_projection_fov} are significantly smaller. This indicates that the HRTF vectors are better represented within the subspace spanned by the steering matrix when attention is confined to the FoV region, making the problem more linearly solvable in that zone.

While this projection analysis uses a hard cutoff outside the FoV, it helps motivate the soft FoV weighting strategy used in practice, highlighting the effectiveness of incorporating FoV weights to improve binaural reproduction accuracy.

\begin{figure}[t]
    \centering
    \includegraphics[trim={45pt 0 0 0},clip,width=0.45\textwidth]{figs/FOV/NullSpace_FOV.jpg}
    \caption{Null space projection values, as in Eq.~(\ref{null_measure}), of the near-field HRTF onto the near-field steering matrix, evaluated across different source distances using FoV weighting.}
    \label{fig:null_projection_fov}
\end{figure}
\fi

\section{Listening Experiment}

This section presents the listening experiment conducted to evaluate the perceptual performance of the proposed BSM methods. The chapter is structured as follows: first, the experimental setup is described, followed by the listening test methodology, and finally, the results are reported.

\subsection{Listening Test Setup}

The listening test setup \footnote{The listening study was conducted at Ben-Gurion University of the Negev.} was largely based on the acoustic simulation framework described in Section~\ref{sec:setup}, using the same source modeling and simulated HRTFs. In this subsection, we highlight only the elements specific to the listening experiment, including the source–positioning, head orientations, and the signal rendering methods used for the test.

The simulated acoustic environment was modeled as a rectangular room with dimensions $[10 \times 6 \times 3]~\text{m}$, using the image source method~\cite{allen1979image}, implemented in MATLAB 2024b. The room was configured with a reverberation coefficient of $R = 0.7$ leading to a reverberation time of $\sim 0.3$ s.

For the listening test, a single omnidirectional point source was placed at coordinates $(x, y, z) = (5,~3.7,~1.7)~\text{m}$, playing a speech signal. The microphone array signals $\mathbf{x}(k)$ were computed using the ATFs described in Section~\ref{sec:setup}.

Two source–listener configurations were evaluated, for which the center of the microphone array was positioned at source distances of $r_s = 0.15~\text{m}$ and $r_s = 0.45~\text{m}$, located at an azimuth of $30^\circ$ to the left of the source and elevation $0^\circ$. Two listener head orientations were tested: one with no head rotation, such that the listener head is aligned with the head of the recording scene, and one with a $40^\circ$ rotation of the listener head to the right. For the rotated case, the source azimuth was adjusted to $-10^\circ$ to maintain the same relative source direction of $30^\circ$ to the right after rotation. The HRTFs used for binaural rendering were also obtained from the same simulation framework.  

BSM filter coefficients were computed using four methods detailed in Secs.~\ref{NF BSM} and \ref{FOV Design}
: (1) Far-Field BSM (FF-BSM), (2) Near-Field BSM (NF-BSM), (3) FF-BSM with FoV weighting (FF-FoV-BSM), and (4) NF-BSM with FoV weighting (NF-FoV-BSM). All filters used the MagLS criterion for frequencies above 1.5~kHz, assuming a high SNR of $+20$ db~ as detailed in sec. \ref{sec:setup}.

As a reference, an additional binaural signal was rendered using high-order Ambisonics (order 14) with simulated HRTFs under the same room conditions. Participants listened to all stimuli over AKG K702 open-back reference headphones, with headphone equalization applied.

\subsection{Methodology}

The listening tests followed the MUltiple Stimuli with Hidden Reference and Anchor (MUSHRA) protocol~\cite{series2014method}. Participants were asked to rate the similarity of each stimulus to the reference, with respect to \emph{Overall Quality}. The experiment consisted of four MUSHRA screens, corresponding to two source distances ($r_s = 0.15~\text{m}$ and $r_s = 0.45~\text{m}$) and two head orientations (with and without $40^\circ$ head rotation). Each screen contained five signals: a hidden reference, and four BSM-based binaural renderings (FF-BSM, NF-BSM, FF-FoV-BSM, and NF-FoV-BSM). While there was not a formal anchor signal in this experiment, the FF-BSM method represents a low quality baseline, as it is not specifically designed for near field sources and does not apply any directional weighting.

14 participants (12 male and 2 female), aged 24 to 37 years, took part in the study. All participants reported normal hearing and were familiar with spatial audio or binaural listening tests.

The experiment was conducted in a quiet environment and consisted of two phases prior to the main evaluation: a training phase and a familiarization phase. In the training phase, participants were introduced to the test interface, headphones, and rating scale. In the familiarization phase, they were allowed to freely explore and listen to all stimuli before starting the formal test.
Participants were encouraged to use the full rating scale and to rely on their perceptual impression of spatial realism, coloration, and overall fidelity. No time limit was imposed, and participants could replay signals as needed before submitting their scores.

\subsection{Results}
\label{sec:results}

\begin{figure*}[!htbp]
    \centering
    % ---- Top row ----
     \begin{subfigure}[b]{0.49\textwidth}
        \centering
        \includegraphics[width=\linewidth]{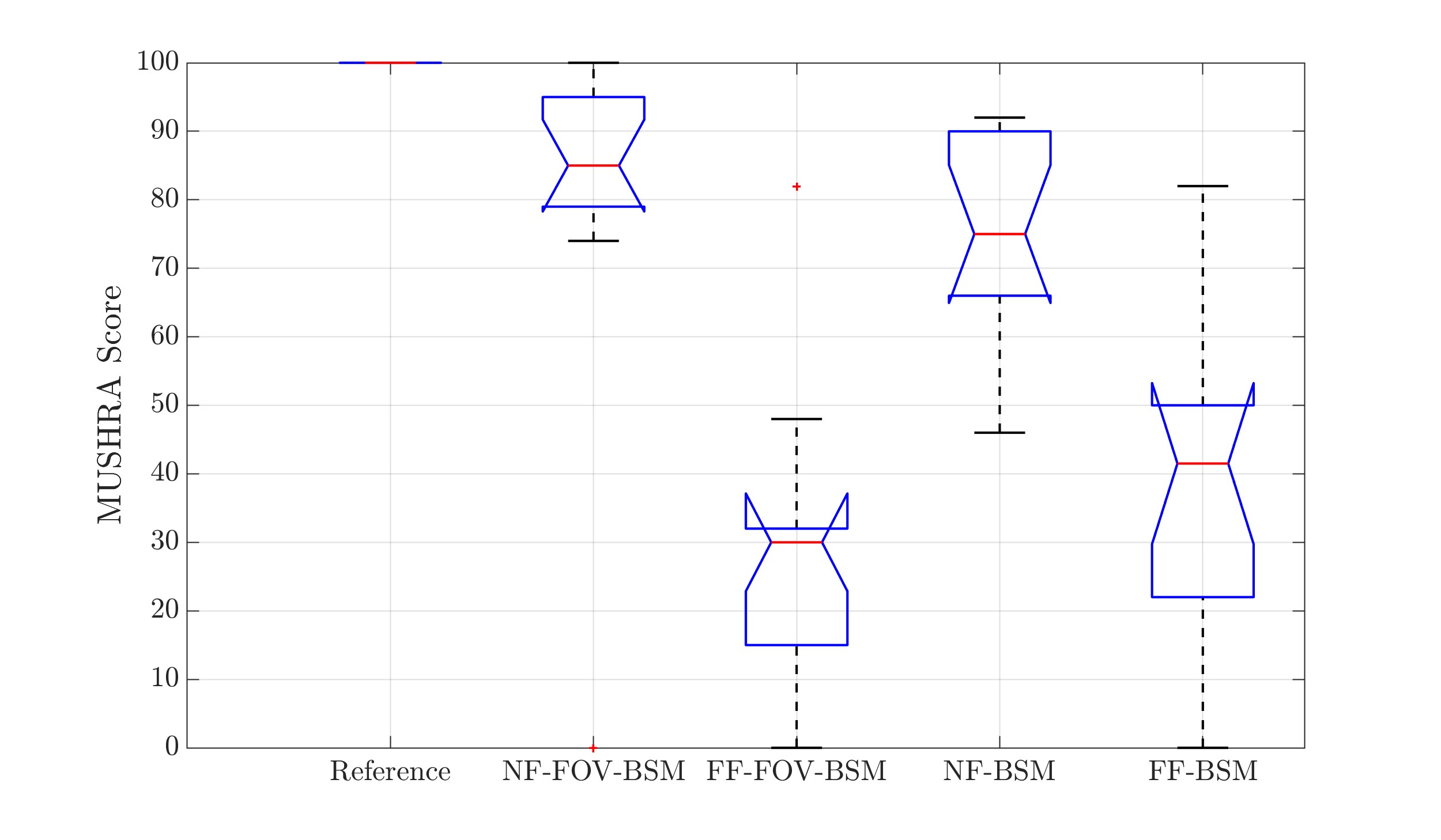}
        \caption{0.15 m – 0°}
        \label{fig:boxplot_015_0}
    \end{subfigure}
    \begin{subfigure}[b]{0.49\textwidth}
        \centering
        \includegraphics[width=\linewidth]{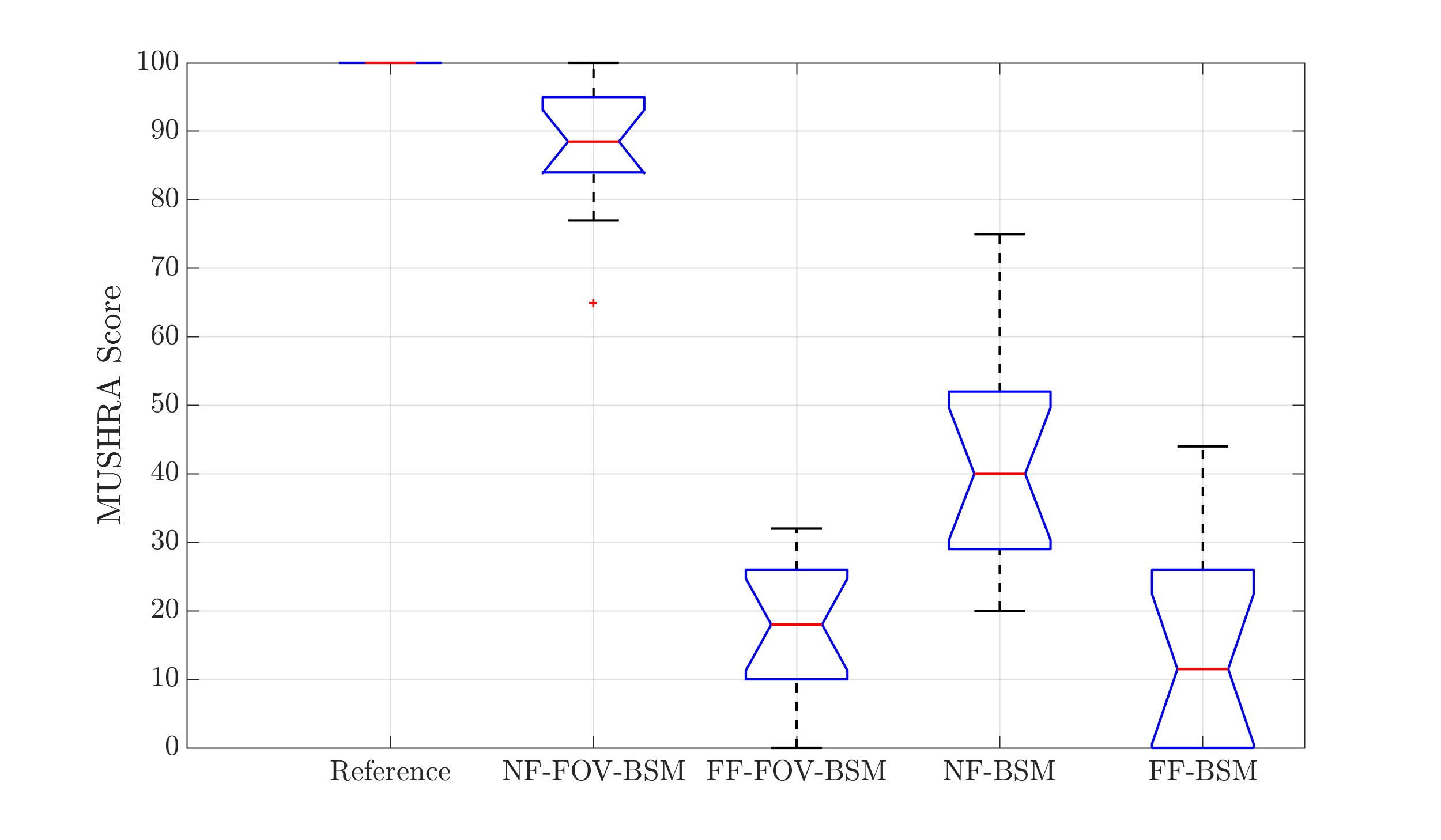}
        \caption{0.15 m – 40°}
        \label{fig:boxplot_015_40}
    \end{subfigure}\hspace{-0.3cm} % negative space to pull columns closer

    % ---- Bottom row ----
    \begin{subfigure}[b]{0.49\textwidth}
        \centering
        \includegraphics[width=\linewidth]{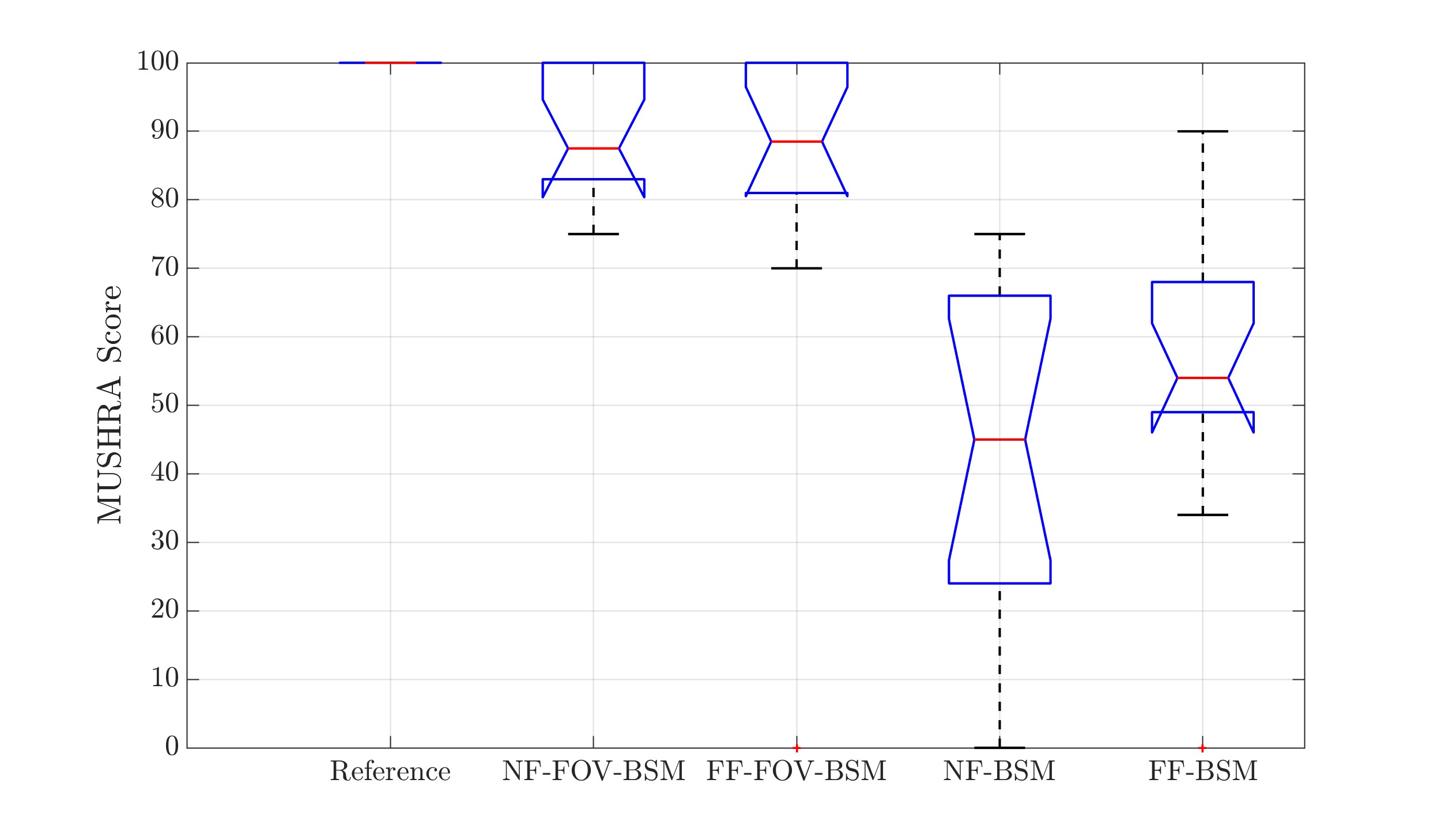}
        \caption{0.45 m – 0°}
        \label{fig:boxplot_045_0}
    \end{subfigure}
    \begin{subfigure}[b]{0.49\textwidth}
        \centering
        \includegraphics[width=\linewidth]{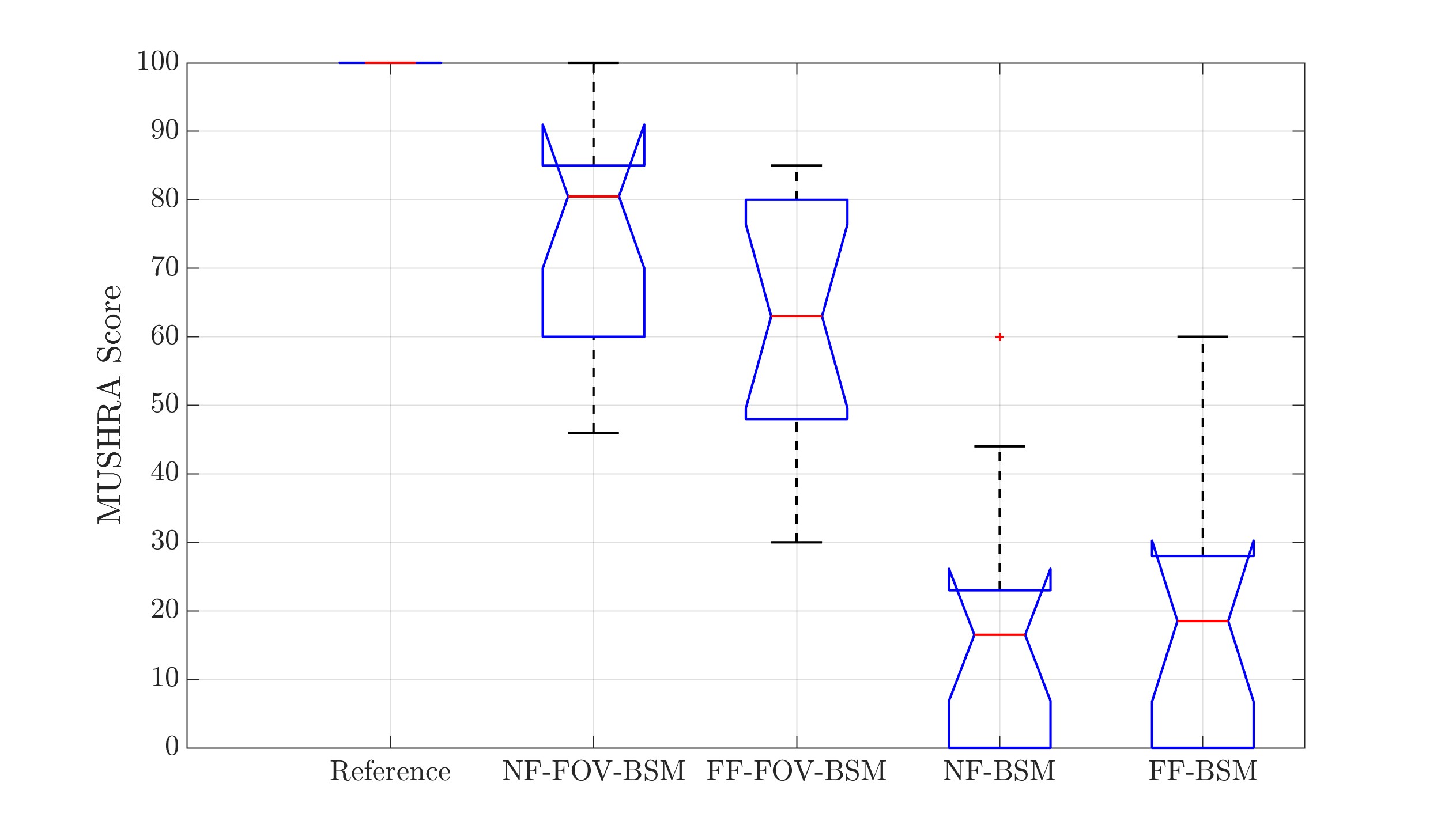}
        \caption{0.45 m – 40°}
        \label{fig:boxplot_045_40}
    \end{subfigure}\hspace{-0.3cm} % negative space to pull columns closer

    \caption{Boxplots of MUSHRA scores for each distance and head orientation combination. Each subplot shows ratings for Reference (High Order Ambisonics), NF-FOV-BSM, NF-BSM, FF-FOV-BSM, and FF-BSM. The boxes represent the interquartile range (IQR) with a line indicating the median. Tukey-style whiskers extend up to 1.5~$\times$~IQR beyond the box, and notches indicate the 95\% confidence interval of the median.
}
    \label{fig:boxplots}
\end{figure*}

The participants’ MUSHRA scores were analyzed using a repeated-measures Analysis of Variance (RM-ANOVA) \cite{keselman2001analysis} with three within-subject factors: \textit{Method} (Reference, NF-FOV-BSM, NF-BSM, FF-FOV-BSM, FF-BSM), \textit{Source Distance} (0.15\,m, 0.45\,m), and \textit{Listener Head Orientation} (0\textdegree, 40\textdegree). The distributions of ratings for each distance/orientation combination are illustrated in Fig.~\ref{fig:boxplots}, which presents four boxplots representing the four combinations of source distance and head orientation.

The RM-ANOVA revealed statistically significant main effects of all three factors:
\begin{itemize}
    \item \textbf{Method}: $F(4, 52) = 72.23$, $p < .001$, $\eta^2 = .967$;
    \item \textbf{Source Distance}: $F(1, 13) = 30.09$, $p < .001$, $\eta^2 = .698$;
    \item \textbf{Head Orientation}: $F(1, 13) = 29.41$, $p < .001$, $\eta^2 = .693$.
\end{itemize}

These main effects indicate that listener ratings varied significantly across methods, differed between source distances, and decreased when the head was rotated by 40\textdegree.

Significant interaction effects were also found:
\begin{itemize}
    \item \textbf{Method $\times$ Distance}: $F(4, 52) = 89.45$, $p < .001$, $\eta^2 = .973$;
    \item \textbf{Method $\times$ Orientation}: $F(4, 52) = 21.16$, $p < .001$, $\eta^2 = .894$;
    \item \textbf{Distance $\times$ Orientation $\times$ Method}: $F(4, 52) = 5.28$, $p = .015$, $\eta^2 = .679$.
\end{itemize}

These interactions demonstrate that the relative ranking of methods depended on both source distance and head orientation.
Pairwise comparisons with Bonferroni correction revealed clear differences between methods. Reference scored significantly higher than all BSM methods, including NF-FOV-BSM ($p < .001$, mean difference = 17.4 points), NF-BSM ($p < .001$, mean difference = 55.5 points), FF-FOV-BSM ($p < .001$, mean difference = 52.3 points), and FF- BSM ($p < .001$, mean difference = 67.8 points).
Among the BSM methods, NF-FOV-BSM achieved the highest ratings. It scored significantly higher than FF-BSM ($p < .001$, mean difference = 50.4 points), FF-FOV-BSM ($p < .001$, mean difference = 34.9 points), and NF-BSM ($p < .001$, mean difference = 38.1 points).
FF-BSM (the traditional BSM method) consistently received the lowest ratings. The differences between FF-BSM and NF-FOV-BSM ($p < .001$) and Reference ($p < .001$) were very large, while the differences with NF-BSM ($p = .024$) and FF-FOV-BSM ($p = .010$) were smaller but still statistically significant.
These results align with the objective findings presented in the simulation study in Section~\ref{Results}, which demonstrated improvements of NF-BSM over FF-BSM, particularly for near-field sources at close distances. Notably, for the closest source distance (0.15\,m), the perceptual gap between NF-BSM and FF-BSM was more pronounced than anticipated from the simulations in Section~\ref{Results}, suggesting that listener perception accentuated the benefits of near-field modeling at very close ranges. In particular, the inclusion of the field-of-view weighting (NF-FOV-BSM) further enhanced performance, as demonstrated in Sec.~\ref{FoV}, resulting in ratings that were closest to the Reference signal. Consistent with the simulations, NF-FOV-BSM was shown to be the most robust method for very close source distances (0.15\,m) and across head orientations.

\section{Conclusions}

This work investigated the performance of BSM methods for near-field sources using realistic simulated data and perceptual evaluation. Building upon our previous study on near-field BSM, this paper extended the analysis to include listener head rotations, evaluation of binaural cues (ILD and ITD), and a newly introduced FoV weighting strategy. 

Results from both the simulation study and the listening experiment consistently showed that NF-BSM outperforms traditional FF-BSM, particularly for sources close to the array, where distance-dependent effects are significant. While the inclusion of source-distance information in NF-BSM reduced binaural reproduction errors, performance degradation remains evident for very close sources, especially under head rotation, highlighting the inherent limitations of wearable arrays in such extreme conditions.

A key contribution of this work is the introduction of FoV weighting for near-field scenarios. This strategy improved robustness and perceptual quality by prioritizing directions of interest. When combined with NF-BSM, the resulting NF-FoV-BSM method substantially outperformed other BSM variants and provided reproduction quality closest to the reference among the evaluated methods.

These findings underscore the limitations of BSM methods that assume far-field plane waves when applied to near-field sources. At the same time, they demonstrate that by incorporating source-distance information and directional weightning, these limitations can largely be overcome, enabling accurate and robust binaural reproduction even for challenging near-field conditions. 

Future work may involve incorporating deep neural networks to enhance binaural reproduction in near-field scenarios, potentially without explicit estimation of source distance or direction. Extending the analysis beyond point sources to more realistic models, such as vibrating pistons that better approximate loudspeakers or human talkers, could provide deeper insight into the physical limitations of current methods. Further investigation into the underlying causes of degradation in near-field reproduction—particularly under extreme proximity or head rotation—may also guide the development of improved algorithms and array designs.



\begin{thebibliography}{10}
\providecommand{\url}[1]{#1}
\csname url@samestyle\endcsname
\providecommand{\newblock}{\relax}
\providecommand{\bibinfo}[2]{#2}
\providecommand{\BIBentrySTDinterwordspacing}{\spaceskip=0pt\relax}
\providecommand{\BIBentryALTinterwordstretchfactor}{4}
\providecommand{\BIBentryALTinterwordspacing}{\spaceskip=\fontdimen2\font plus
\BIBentryALTinterwordstretchfactor\fontdimen3\font minus \fontdimen4\font\relax}
\providecommand{\BIBforeignlanguage}[2]{{%
\expandafter\ifx\csname l@#1\endcsname\relax
\typeout{** WARNING: IEEEtran.bst: No hyphenation pattern has been}%
\typeout{** loaded for the language `#1'. Using the pattern for}%
\typeout{** the default language instead.}%
\else
\language=\csname l@#1\endcsname
\fi
#2}}
\providecommand{\BIBdecl}{\relax}
\BIBdecl

\bibitem{madmoni2018direction}
L.~Madmoni and B.~Rafaely, ``Direction of arrival estimation for reverberant speech based on enhanced decomposition of the direct sound,'' \emph{IEEE Journal of Selected Topics in Signal Processing}, vol.~13, no.~1, pp. 131--142, 2018.

\bibitem{rafaely2022spatial}
B.~Rafaely, V.~Tourbabin, E.~Habets, Z.~Ben-Hur, H.~Lee, H.~Gamper, L.~Arbel, L.~Birnie, T.~Abhayapala, and P.~Samarasinghe, ``Spatial audio signal processing for binaural reproduction of recorded acoustic scenes--review and challenges,'' \emph{Acta Acustica}, vol.~6, p.~47, 2022.

\bibitem{richard2023audio}
G.~Richard, P.~Smaragdis, S.~Gannot, P.~A. Naylor, S.~Makino, W.~Kellermann, and A.~Sugiyama, ``Audio signal processing in the 21st century: The important outcomes of the past 25 years,'' \emph{IEEE Signal Processing Magazine}, vol.~40, no.~5, pp. 12--26, 2023.

\bibitem{ben2017spectral}
Z.~Ben-Hur, F.~Brinkmann, J.~Sheaffer, S.~Weinzierl, and B.~Rafaely, ``Spectral equalization in binaural signals represented by order-truncated spherical harmonics,'' \emph{The Journal of the Acoustical Society of America}, vol. 141, no.~6, pp. 4087--4096, 2017.

\bibitem{madmoni2025design}
L.~Madmoni, Z.~Ben-Hur, J.~Donley, V.~Tourbabin, and B.~Rafaely, ``Design and analysis of binaural signal matching with arbitrary microphone arrays and listener head rotations,'' \emph{EURASIP Journal on Audio, Speech, and Music Processing}, vol. 2025, no.~1, p.~11, 2025.

\bibitem{rafaely2015fundamentals}
B.~Rafaely, \emph{Fundamentals of spherical array processing}.\hskip 1em plus 0.5em minus 0.4em\relax Springer, 2015, vol.~8.

\bibitem{gerzon1985ambisonics}
M.~A. Gerzon, ``Ambisonics in multichannel broadcasting and video,'' \emph{Journal of the Audio Engineering Society}, vol.~33, no.~11, pp. 859--871, 1985.

\bibitem{song2008using}
W.~Song, W.~Ellermeier, and J.~Hald, ``Using beamforming and binaural synthesis for the psychoacoustical evaluation of target sources in noise,'' \emph{The Journal of the Acoustical Society of America}, vol. 123, no.~2, pp. 910--924, 2008.

\bibitem{ifergan2022selection}
I.~Ifergan and B.~Rafaely, ``On the selection of the number of beamformers in beamforming-based binaural reproduction,'' \emph{EURASIP Journal on Audio, Speech, and Music Processing}, vol. 2022, no.~1, p.~6, 2022.

\bibitem{madmoni2024design}
L.~Madmoni, Z.~Ben-Hur, J.~Donley, V.~Tourbabin, and B.~Rafaely, ``Design and analysis of binaural signal matching with arbitrary microphone arrays,'' \emph{arXiv preprint arXiv:2408.03581}, 2024.

\bibitem{deppisch2021end}
T.~Deppisch, H.~Helmholz, and J.~Ahrens, ``End-to-end magnitude least squares binaural rendering of spherical microphone array signals,'' in \emph{2021 Immersive and 3D Audio: from Architecture to Automotive (I3DA)}.\hskip 1em plus 0.5em minus 0.4em\relax IEEE, 2021, pp. 1--7.

\bibitem{ahrens2021spherical}
J.~Ahrens, H.~Helmholz, D.~L. Alon, and S.~V. Amengual~Gar{\'\i}, ``Spherical harmonic decomposition of a sound field based on observations along the equator of a rigid spherical scatterer,'' \emph{The Journal of the Acoustical Society of America}, vol. 150, no.~2, pp. 805--815, 2021.

\bibitem{mccormack2022parametric}
L.~McCormack, A.~Politis, R.~Gonzalez, T.~Lokki, and V.~Pulkki, ``Parametric ambisonic encoding of arbitrary microphone arrays,'' \emph{IEEE/ACM Transactions on Audio, Speech, and Language Processing}, vol.~30, pp. 2062--2075, 2022.

\bibitem{fisher2010near}
E.~Fisher and B.~Rafaely, ``Near-field spherical microphone array processing with radial filtering,'' \emph{IEEE Transactions on Audio, Speech, and Language Processing}, vol.~19, no.~2, pp. 256--265, 2010.

\bibitem{goldring2025binauralsignalmatchingwearable}
\BIBentryALTinterwordspacing
S.~Goldring, Z.~B. Hur, D.~L. Alon, and B.~Rafaely, ``Binaural signal matching with wearable arrays for near-field sources,'' 2025. [Online]. Available: \url{https://arxiv.org/abs/2507.15517}
\BIBentrySTDinterwordspacing

\bibitem{kan2009psychophysical}
A.~Kan, C.~Jin, and A.~van Schaik, ``A psychophysical evaluation of near-field head-related transfer functions synthesized using a distance variation function,'' \emph{The Journal of the Acoustical Society of America}, vol. 125, no.~4, pp. 2233--2242, 2009.

\bibitem{brungart1999auditory}
D.~S. Brungart and W.~M. Rabinowitz, ``Auditory localization of nearby sources. head-related transfer functions,'' \emph{The Journal of the Acoustical Society of America}, vol. 106, no.~3, pp. 1465--1479, 1999.

\bibitem{brungart2002near}
D.~S. Brungart, ``Near-field virtual audio displays,'' \emph{Presence}, vol.~11, no.~1, pp. 93--106, 2002.

\bibitem{xie2013head}
B.~Xie, \emph{Head-related transfer function and virtual auditory display}.\hskip 1em plus 0.5em minus 0.4em\relax J. Ross Publishing, 2013.

\bibitem{slaney1998auditory}
M.~Slaney, ``Auditory toolbox,'' \emph{Interval Research Corporation, Tech. Rep}, vol.~10, no. 1998, p. 1194, 1998.

\bibitem{katz2014comparative}
B.~F. Katz and M.~Noisternig, ``A comparative study of interaural time delay estimation methods,'' \emph{The Journal of the Acoustical Society of America}, vol. 135, no.~6, pp. 3530--3540, 2014.

\bibitem{gayer2024ambisonics}
Y.~Gayer, V.~Tourbabin, Z.~Ben-Hur, J.~Donley, and B.~Rafaely, ``Ambisonics encoding for arbitrary microphone arrays incorporating residual channels for binaural reproduction,'' \emph{arXiv preprint arXiv:2402.17362}, 2024.

\bibitem{gomez2020pinna}
J.~G{\'o}mez~Bola{\~n}os, M.~Geronazzo, R.~Mehra, L.~Savioja \emph{et~al.}, ``Pinna-related transfer functions and lossless wave equation using finite-difference methods: Validation with measurements,'' \emph{The Journal of the Acoustical Society of America}, vol. 147, no.~5, pp. 3631--3645, 2020.

\bibitem{hajarolasvadi2024effect}
S.~Hajarolasvadi, M.~Khaleghimeybodi, P.~Razavi, M.~Smirnov, and S.~T. Prepeli{\c{t}}{\u{a}}, ``Effect of sound-induced vibrations of the pinna on head-related transfer functions: Experimental and numerical investigations,'' \emph{The Journal of the Acoustical Society of America}, vol. 155, no.~4, pp. 2875--2890, 2024.

\bibitem{oreinos2013measurement}
C.~Oreinos and J.~M. Buchholz, ``Measurement of a full 3d set of hrtfs for in-ear and hearing aid microphones on a head and torso simulator (hats),'' \emph{Acta Acustica united with Acustica}, vol.~99, no.~5, pp. 836--844, 2013.

\bibitem{blauert1997spatial}
J.~Blauert, \emph{Spatial hearing: the psychophysics of human sound localization}.\hskip 1em plus 0.5em minus 0.4em\relax MIT press, 1997.

\bibitem{yost1988discrimination}
W.~A. Yost and R.~H. Dye~Jr, ``Discrimination of interaural differences of level as a function of frequency,'' \emph{The Journal of the Acoustical Society of America}, vol.~83, no.~5, pp. 1846--1851, 1988.

\bibitem{mossop1998lateralization}
J.~E. Mossop and J.~F. Culling, ``Lateralization of large interaural delays,'' \emph{The Journal of the Acoustical Society of America}, vol. 104, no.~3, pp. 1574--1579, 1998.

\bibitem{andreopoulou2017identification}
A.~Andreopoulou and B.~F. Katz, ``Identification of perceptually relevant methods of inter-aural time difference estimation,'' \emph{The Journal of the Acoustical Society of America}, vol. 142, no.~2, pp. 588--598, 2017.

\bibitem{allen1979image}
J.~B. Allen and D.~A. Berkley, ``Image method for efficiently simulating small-room acoustics,'' \emph{The Journal of the Acoustical Society of America}, vol.~65, no.~4, pp. 943--950, 1979.

\bibitem{series2014method}
B.~Series, ``Method for the subjective assessment of intermediate quality level of audio systems,'' \emph{International Telecommunication Union Radiocommunication Assembly}, vol.~2, 2014.

\bibitem{keselman2001analysis}
H.~Keselman, J.~Algina, and R.~K. Kowalchuk, ``The analysis of repeated measures designs: a review,'' \emph{British Journal of Mathematical and Statistical Psychology}, vol.~54, no.~1, pp. 1--20, 2001.

\end{thebibliography}
\end{document}